\documentclass[structabstract]{aa}  
\usepackage{txfonts}
\usepackage{graphicx}
\usepackage{natbib}
\usepackage{aalongtable}
\bibpunct[; ]{(}{)}{;}{a}{,}{,}
\begin{document}
   \title{Determination of stellar parameters of C-rich hydrostatic stars from spectro-interferometric observations}


   \author{C. Paladini\inst{1},
     G.~T. van Belle\inst{2},
     B. Aringer\inst{3},
     J.Hron\inst{1},
     P. Reegen\inst{1}\fnmsep\thanks{Deceased 5 February 2011.},
     C.~J. Davis\inst{4},
     \and    
     T. Lebzelter\inst{1}
              }

   \institute{Institute for Astronomy (IfA), University of Vienna,
              T\"urkenschanzstrasse 17, A-1180 Vienna\\
              \email{claudia.paladini@univie.ac.at}
         \and
         European Southern Observatory, Karl-Schwarzschild-Str. 2 D-85748 Garching bei M\"unchen, Germany
         \and
         INAF-OAPD, Vicolo dell'Osservatorio 5, 35122 Padova, Italy
         \and
       Joint Astronomy Centre, 660 North A'oh\={o}k\={u} Place, University Park, Hilo, Hawaii 96720, USA \\
   }

   \date{Received; accepted}

 
  \abstract
   {Giant stars, and especially C-rich giants, contribute significantly to the chemical enrichment of galaxies.
    The determination of precise parameters for these stars 
    is a necessary prerequisite for a proper implementation of this evolutionary phase
    in the models of galaxies.   
    Infrared interferometry opened new horizons in the study of the stellar parameters of giant stars, and
    provided new important constraints for the atmospheric and evolutionary models.}
   {We aim to determine which stellar parameters can be constrained by using infrared interferometry
     and spectroscopy, in the case of C-stars what is the precision which can be achieved and what are the limitations.}
   {For this purpose we obtained new infrared spectra and combined them with unpublished interferometric measurements for 
     five mildly variable carbon-rich asymptotic giant branch stars. The observations were compared with 
     a large grid of hydrostatic model atmospheres and with new isochrones which include the predictions of the thermally pulsing phase.}
   {For the very first time we are able to reproduce spectra in the range between 0.9 and 4 $\mu$m, and $K$ broad band interferometry
     with hydrostatic model atmospheres. 
     Temperature, mass, log$(g)$, C/O and a reasonable range for the distance were derived for all the objects of our study.
   All our targets have at least one combination of best-fitting parameters which lays in the region of the HR-diagram where C-stars are predicted.}
   {We confirm that low resolution spectroscopy is not sensitive to the mass and log$(g)$ determination. 
     For hydrostatic objects the $3\,\mu$m feature is very sensitive to temperature variations therefore it is a very powerful tool for accurate temperature
     determinations. Interferometry can constrain mass, radius and log$(g)$ but a distance has to be assumed. The large uncertainty in the distance
     measurements available for C-rich stars remains a major problem.}

\authorrunning{Paladini et al.}
   \titlerunning{Determination of stellar parameters of C-rich hydrostatic stars}
   
   \maketitle
%

\section{Introduction}

The basic properties and evolutionary status of a star
can be determined knowing its mass, luminosity, radius and chemical composition.
Nowadays we are reaching a point where mass estimates 
are available not only for binary but also for single objects \citep[see the recent review by][]{auf09}.
This major advance has been mainly achieved by establishing interferometry
as a standard tool for investigating stars \citep{wit04a}.
The combination of high angular and spectral resolution 
gives the possibility to study the spatial structure of
single stellar objects, which were before treated only as point sources.
Red giant stars are very good targets for interferometric investigations because
of their extended atmosphere and their brightness in the infrared. 
These stars are main contributors to the infrared light
and to the chemical enrichment of galaxies. 
Therefore an accurate determination of their fundamental parameters
is mandatory for a proper implementation in the models of galaxies.

\citet{wit01} made a first attempt to derive the parameters of a sample of asymptotic giant branch (AGB) stars, 
by comparing the spectro-interferometric observations with Kurucz
model atmosphere. The uncertainty in parallax and bolometric flux dominate the errors.
As a recent example, \citet{nei08} determined the fundamental parameters of three M-type stars. 
They combined infrared interferometry with spectro-photometric observations.
The interferometric data were obtained with the VLTI/VINCI instrument, and
 previously interpreted by using {\sc{Phoenix}} and {\sc{Atlas}} atmospheric models
\citep[][respectively]{hau99, kur93} in the work of \citet{wit04b, wit06a, wit06b}.
\citet{nei08} used the new generation of {\sc{SAtlas}}
models \citep{les08} for the interpretation of the data: in this way the authors were able not 
only to determine the fundamental parameters,
but also to constrain the atmospheric models used.
The typical approach followed in these works to determine the stellar parameters is:
(i) determination of the limb darkened radius by fitting the interferometric observations, the resulting radius is converted 
in Rosseland radius; (ii) a linear radius is derived by assuming a distance; (iii) an effective temperature is derived 
by fitting the SED; (iv) the luminosity is obtained using radius and $T_{\rm{eff}}$; (v) the gravity and the mass of the star are 
constrained by comparison with the theoretical models.  

The targets of our study are 
an important subclass of the AGB objects: the carbon-rich AGB stars. According to theory only stars with a range of mass
between 1 and 4 $M_{\sun}$ undergo the third dredge-up during their AGB phase,
with the result of increasing their C/O ratio \citep{ibe83}. 
The spectra of these giants are consequently dominated by the absorption features of
carbonaceous molecular species like: CO, CN, HCN, C$_2$, C$_3$, C$_2$H$_2$ \citep{tsu86, jor00, loi01}.
As a result, these objects are main contributors to the C-enrichment of the ISM.
  
The state-of-the-art model atmospheres for C-stars are very promising, being able 
to produce large grids of models \citep{ari09, mat09} 
which describe the observed properties of these objects reasonably well.
The hydrostatic model atmospheres for C-stars were compared with optical and infrared spectroscopy in 
\citet{jor00, loi01, abi10} and references therein. 
The major difference between an earlier generation of models as used by \citep{jor00,loi01} and the hydrostatic atmospheres
applied in our work \citep{ari09}
is the inclusion of atomic opacities and an update to new and more accurate molecular opacities.

The stellar evolutionary calculations reached a highly sophisticated level as well, being able to include 
a detailed thermally pulsing phase with third dredge-up, hot-bottom
burning and variable molecular opacity \citep{mar08}.

The purpose of this work is to find out which stellar parameters can be determined
for weakly variable C-stars (visual amplitude $\le 2$ mag) by using spectroscopy, infrared interferometry and hydrostatic model atmospheres; 
and what is the accuracy.
We aim to test the stellar atmospheric models with different techniques and a multiwavelength approach.
This is a very important step since errors in the model structure
limit quantitative checks of stellar evolution.
Our approach in the parameter determination is slightly different from the one adopted by 
\cite{nei08} for a few simple reasons: (1) the Rosseland radius is not a direct observable, 
and in the C-stars (to the contrary of M-type AGB stars) 
there are no windows in the near infrared where to measure a ``continuum'' radius \citep{pal09}; 
(2) it is very rare to obtain simultaneous multiwavelength photometric observations, which is mandatory when 
dealing with variable stars. Nevertheless the parameters determined in this work will be compared in a follow-up paper 
(van~Belle et al., in prep.) with the one obtained
by combining photometric and interferometric observations.
  
A short description of the observations acquired and of the data reduction
for both spectroscopy and interferometry will be given in Sect.~\ref{obs.sect}. The hydrostatic model atmospheres and the methods to determine
the observables will be presented in Sect.~\ref{mod.sect}. The approach used to determine the stellar parameters (temperature, 
C/O, mass, log$(g)$) is presented in Sect.~\ref{pardet.sect}, while in Sect.~\ref{evol.sect} we summarise the comparison with the 
evolutionary tracks. We present a detailed discussion of the single objects studied in this work in Sect.~\ref{obj.sect}, 
followed by a more general discussion on the obtained results (Sect.~\ref{disc.sect}) and the conclusions (Sect.~\ref{concl.sect}).


   \section{Observations and data reduction}
   \label{obs.sect}
   Five mildly variable stars were selected for this investigation: \object{CR Gem}, \object{HK Lyr},
   \object{RV Mon}, \object{Z Psc}, \object{DR Ser}.
   In Table~\ref{obs.tab} the coordinates of the objects, near-infrared photometry, and the $12\,\mu$m IRAS flux 
   are presented together with period (P) and amplitude (A) of variability.
   Unless otherwise stated A and P are derived from the General Catalogue of Variable Stars \citep[GCVS;][]{sam09}.
   The amplitude values list the maximum amplitude recorded, and might be based only on few data points, sometimes recorded on photographic plates. 
   More recent observations from the public surveys such as ASAS \citep{poj02}, and Hipparcos Variability Annex \citep{bei88}, show
   smaller $V$ amplitude variations, often quite stable over a long period. These more recent values are listed in the discussion on the individual targets
   (Sect.~\ref{obj.sect}).
   The objects selected are semiregular or irregular variables.  
  The variability amplitude, and the period
     indicate that the atmospheres of our targets can be represented with hydrostatic models.
     The literature values of A and P are small in comparison to dynamic objects such as Mira stars 
     (P longer than one year, and A of several $V$ magnitudes). 
     In the recent grid of dynamic models for C-stars from \cite{mat09} almost all the models with 
     a period lower than 250 days do not develop a stellar wind.
     The few exceptions are models with extremely low temperature, or very high C/O value.
     The same applies to the predictions of the previous generations of dynamic models \citep[i.e. Table~1][]{gau04}.
     The temperature-density structure of these windless models is only slightly varying around the 
     hydrostatic configuration. These small variations of the atmospheric structure are also reflected
     in the resulting spectra,but as can be seen in Fig.\,A.1 of \cite{now10}, the spectra of objects
     with very mild pulsations and no mass loss do not differ significantly from the corresponding one
     based on a hydrostatic model.
     The hydrostatic spherical symmetric approximation 
    will be checked with interferometric observations (where possible) also in Sect.~\ref{interf.sect}.
    We collected for all the targets $IJHKL$ infrared spectroscopy, and $K$ broad band
   interferometric observations.
   \begin{table*}[!th]
     \caption{\label{obs.tab}Targets list, available photometry and variability information.}
     \centering
     \begin{tabular}{l l l l l l l l l l l}
       \hline
       \hline
       ID &RA & DEC & $J$   & $H$   & $K$   & $L$ & f$_{[12\,\mu{\rm{m}}]}$& Var. Type & Amplitude & Period\\
          &   &     & [mag] & [mag] &  [mag]&[mag]&   [Jy]              &                   &[mag]     &[days]\\
       \hline
         CR~Gem & 06:34:23.92 &+16:04:30.30  & 3.36 & 2.07 & 1.46 & $0.88$\tablefootmark{a}& $39.31$ &  Lb & 1.20~$B$ & 250 \\
         HK~Lyr & 18:42:50.00 &+36:57:30.89  & 3.23 & 2.15 & 1.62 & $0.99$\tablefootmark{b}& $22.74$ &  Lb & 0.40~$V$\tablefootmark{f} & $186$\tablefootmark{f} \\
         RV~Mon  & 06:58:21.49 &+06:10:01.50 & 3.06 & 1.96 & 1.43 & $0.85$\tablefootmark{c}& $31.13$ &  SRb& 2.19~$B$ & 131 \\
         Z~Psc   & 01:16:05.03 &+06:10:01.50 & 2.11 & 1.10 & 0.70 & $0.41$\tablefootmark{d}& $33.42$ &  SRb& 1.30~$p$ & 144   \\
         DR~Ser & 18 47 21.02 &+05 27 18.60   & 4.12 & 2.79 & 2.08 & $1.38$\tablefootmark{e}&$15.9 $ &  Lb & 0.50~$V$\tablefootmark{g} & $196$\tablefootmark{g}  \\
         \hline
   \end{tabular}
     \tablefoot{The amplitude and the period (unless otherwise stated)
      are derived from the General Catalogue of Variable Stars \citep[GCVS;][]{sam09}. 
       The band used for amplitude of variability measurement is also reported; 
       '$p$' stays for photographic plates.
\tablefoottext{a}{\citet{ker96a}};
\tablefoottext{b}{\citet{ker96b}};
\tablefoottext{c}{\citet{epc90}};
\tablefoottext{d}{\citet{bag96}};
\tablefoottext{f}{\citet{bei88}};
\tablefoottext{g}{\citet{poj02}}.
        }
     \end{table*}

  \subsection{Spectroscopy} 
  \label{spec.sect}
  The spectra for our targets were obtained with the UIST instrument \citep{ram04} on the United Kingdom InfraRed Telescope
   (UKIRT) as part of the UKIRT Service Programme: {\tt{u/serv/1790}} and {\tt{u/serv/1810}}.
   The first run of observations was on 29 June 2008; the second one on 20, 23, 25, and 26 January 2009. 
   Four grism spectra were collected for every target, covering the following spectral ranges: 
   $IJ$ with wavelengths in the interval\footnote{Although the wavelengths range of the $IJ$ grism starts at 0.86 $\mu$m, 
   only the part beyond 0.9 $\mu$m is usable because the spectral blocking filter blocks emission 
   (essentially has zero transmission) below this wavelength
   and, in any case, the instrumental throughput drops considerably towards the $I$ band.} 
   $[0.862, 1.418]$ $\mu$m; $HK$ with wavelengths in the interval $[1.395, 2.506]$ $\mu$m; 
   short$L$, and long$L$ bands covering $[2.905, 3.638]$ and $[3.620, 4.232]$ $\mu$m, respectively.
   We obtained also spectra for three standard stars to be used for the calibration procedure.
   The spectral type, photometry, and effective temperature of the standard stars are listed in Table~\ref{standard.tab}.
\begin{table}[!th]
  \caption{\label{standard.tab}Spectral type, photometry and effective 
      temperature of the standard stars used for the spectroscopic calibration.}   
  \centering
  \begin{tabular}{l l l l l l l}
    \hline
    \hline
    ID & Sp.~Type & $V$ & $J$ & $H$ & $K$ &$T_{\rm{eff}}$ \\
       &          & [mag]& [mag] &[mag] & [mag] &[K]\\
    \hline
    HIP~92946   & A5V & 4.62  & 4.160 & 4.163 & 4.085 & $7\,880$\tablefootmark{a} \\
    HIP~33024 & F0V & 5.753 & 5.308 & 5.232 & 5.088 & $7\,400$\tablefootmark{b}  \\
    HIP~5544 & F0V & 5.160 & 4.764 & 4.545 & 4.393 & $7\,500$\tablefootmark{c}  \\
    \hline
  \end{tabular}
\tablefoot{The photometry is from the 2MASS All Sky catalogue \citep{cut03};
\tablefoottext{a}\citet{ers03};
\tablefoottext{b}\citet{mas06};
\tablefoottext{c}\citet{sol97}.
}
\end{table}   
Since our objects were bright, we observed with a 2-pixel (0.24 arcsec) wide slit.
   Targets and standard stars were nodded along the slit, with spectra being taken in an A-B-B-A 
   sequence to facilitate first-order sky subtraction. 
   Exposure times of 2-coadds $\times$ 2 seconds or 
   1-coadd  $\times$ 4 seconds were used with each grism, depending on the brightness of the source.   
   The nodding sequence was executed twice, resulting in a total integration time of 32 seconds per source in each grism. 
   The resolution of the spectra ranges between 400 ($IJ$) and 1800 (long$L$).   
   \object{HIP 92946} was used to remove telluric features for targets observed in 2008 (HK~Lyr, DR~Ser), while \object{HIP 33024}
   and \object{HIP 5544} were used for the 2009 observations (CR~Gem, RV~Mon and Z~Psc, respectively).
   
   Data reduction was performed using the ORAC-DR UIST pipeline \citep{cav03}.
   The subtraction of the sky removes the OH lines and the thermal background emission. At this stage the 
   spectrum is made of a positive and a negative part which can be extracted. An Argon lamp is used for
   the calibration of the object in wavelength. Before every target observation a telluric standard was observed.
   The observations of target and reference star are carried out always in a range of 
   airmass between 1.0 and 1.3. Only in one case we had airmass 1.5.
   The airmass difference between target and standard is usually of the order of 0.1 thus producing a negligible error in the flux calibration.
   Once that the object and the standard were calibrated in wavelength, the telluric correction was performed.
   For this purpose we used the routine {\tt IRFLUX} from the Starlink\footnote{Starlink software is available for downloading on the webpage 
   {\tt{http://starlink.jach.hawaii.edu/}}.} software package. 
   {\tt IRFLUX} divides the standard spectrum by a black body with assigned temperature (the one of the standard).
   The outcome of this operation is a spectrum including only the telluric features, atmospheric and instrumental effects.
   Finally the target is divided by this last determined spectrum and the telluric calibration is performed.
   The standard stars are hot enough to be approximated with a black body in the covered wavelength range.

\begin{figure}
      \centering
      \includegraphics[angle = 90,width = \hsize]{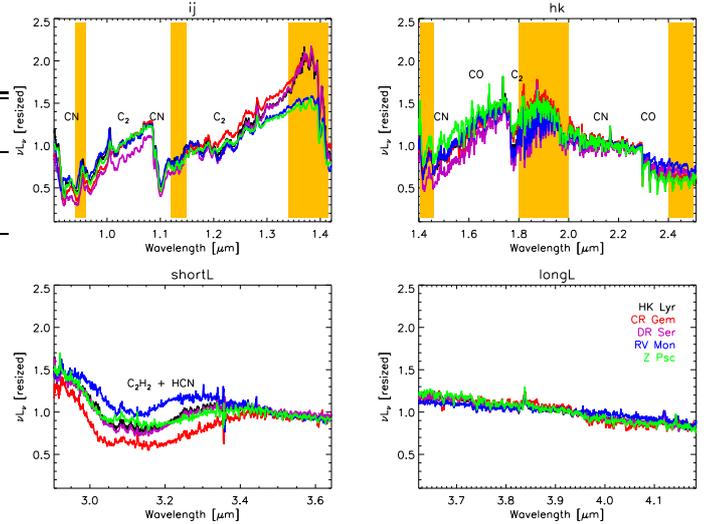}
      \caption{Comparison between all the spectra of our targets. The main molecular features are indicated. The shadowed (yellow)
      regions are affected by absorption of the Earth atmosphere.}
      \label{allspec.fig}
    \end{figure}

   Once the telluric correction was performed a few emission lines appeared in the spectrum of the targets.
   Those are residual of the atmosphere of the standards which in this part of the spectrum does not contribute 
   with continuum (this is flat), but it shows few hydrogen lines. 
   We identified those lines by comparing the spectrum of the standards with the corresponding hydrostatic models from \cite{shu04}.
   The region where the lines are located was not taken into account in the fitting procedure.
   as well as the data at the edge of the atmospheric windows (shaded region in Fig.~\ref{allspec.fig}) 
  
   The reddening correction is always a very delicate task in spectroscopy. 
   Every spectrum was dereddened using the {\tt EXTIN.PRO} IDL procedure \citep{amo05}. 
   The code computes the visible extinction ($A_{\rm V}$) along a path from the
   Sun to any point in the Galaxy, specified by galactic coordinates and
   distance. Unfortunately the distance measurements associated with our objects are very uncertain.
   We selected three reference distances for each target from the literature: the distance 
   measured by the Hipparcos satellite \citep{per97},
   a distance estimated assuming a constant $K$ magnitude \citep{cla87}, and a distance estimated by \citet{ber05}
   which is based on Hipparcos measurements but taking into account three distinct biases \citep[see Sect.~2.3 of][]{ber02}. 
   Hereafter the different distances will be indicate with the abbreviation $d_{\rm{Ber}}$,  $d_{\rm{Cla}}$,  $d_{\rm{Hipp}}$.
   
   The distances for each object are listed in Table~\ref{dist.tab}. 
   The values are in most of the cases very different between each other, but no trend or systematic behaviour 
   can be identified.
   For the Hipparcos measurements
   the errors associated to the distances are also given.
   No error estimation was found in literature for $d_{\rm{Berg}}$ and $d_{\rm{Clau}}$ therefore we assume at least
   $20\%$ of uncertainty. 
   This corresponds to $40\%$ error on the luminosity.

   Although the distances have very different values the resulting reddening correction is very similar (as expected since these
   are nearby objects).
   We tested the effect of the different reddening correction on the estimation of the parameters (in particular the temperature). 
   The typical shift for the temperature determinations is of the order of 10 K,
   which is low compared to the error bar.

   Except for DR~Ser, where no correction was applied (see detailed discussion in Sect.~\ref{drser.sect}), 
   we used for our computation the reddening correction
   estimated for the Hipparcos distance.

   \subsection{Interferometry}
   \label{interf.sect}
  
Observations for this investigation were primarily taken with the Palomar
Testbed Interferometer (PTI).
PTI was an 85 to 110 m  baselines $H$ and $K$ band (1.6 $\mu$m and 2.2 $\mu$m)
interferometer located at the Palomar Observatory in San Diego County,
California,
and is described in detail in \citet{col99}. It had
three 40-cm apertures used
in pairwise combination for measurements of stellar fringe visibility on
sources that range in angular size from 0.05 to 5.0 milliarcseconds, being
able to resolve individual sources $\theta > 1.0$ mas in size.  PTI had been
in nightly operation since 1997 and was decommissioned in 2009, with
minimum downtime during the
intervening years.  In addition to the carbon stars observed as part of
this investigation,
appropriate calibration sources were observed as well and can be found in
\citet{van08}.

The calibration of the squared visibility ($V^2$) data is performed by estimating the
interferometer system visibility ($V^2_{\textrm{\tiny SYS}}$) using the
calibration sources with model angular diameters and then normalising the raw
carbon star visibility by $V^2_{\textrm{\tiny SYS}}$ to estimate the $V^2$
measured by an ideal interferometer at that epoch
\citep{moz91, bod98, van05}.
Uncertainties in the system visibility and the calibrated target visibility
are inferred from internal scatter among the data in an observation using
standard error-propagation calculations \citep{bod99}. Calibrating our
point-like calibration objects against each other produced no evidence of
systematics, with all objects delivering reduced $V^2 = 1$.

PTI's limiting night-to-night measurement error is
$\sigma_{V^2_{\textrm{\tiny
SYS}}}\approx 1.5 -1.8$\%, the source of which is most likely a
combination of
effects: uncharacterized atmospheric seeing (in particular, scintillation),
detector noise, and other instrumental effects. This measurement error limit
is an empirically established floor from the previous study of \citet{bod99}.

   In our sample there is one star, \object{Z Psc}, without PTI observations.
   One visibility measurement taken with IOTA (Infrared Optical Telescope Array) is available in literature 
   for this object. 
   The visibility point value was obtained in the $K$ filter ($\lambda = 2.2$ $\mu$m, $\Delta\lambda = 0.4$ $\mu$m)
   and it is published in Table~1 of \citet{dyc96}.
   We will use this value for our investigation.
   
   Table~\ref{vis.tab} gives an overview on the observational parameters
   of the interferometric observations used, namely 
   UT date, baseline, position angle, calibrated visibility with associated error.
 
\begin{figure*}
           \centering
           \includegraphics[angle = 90,width = \hsize]{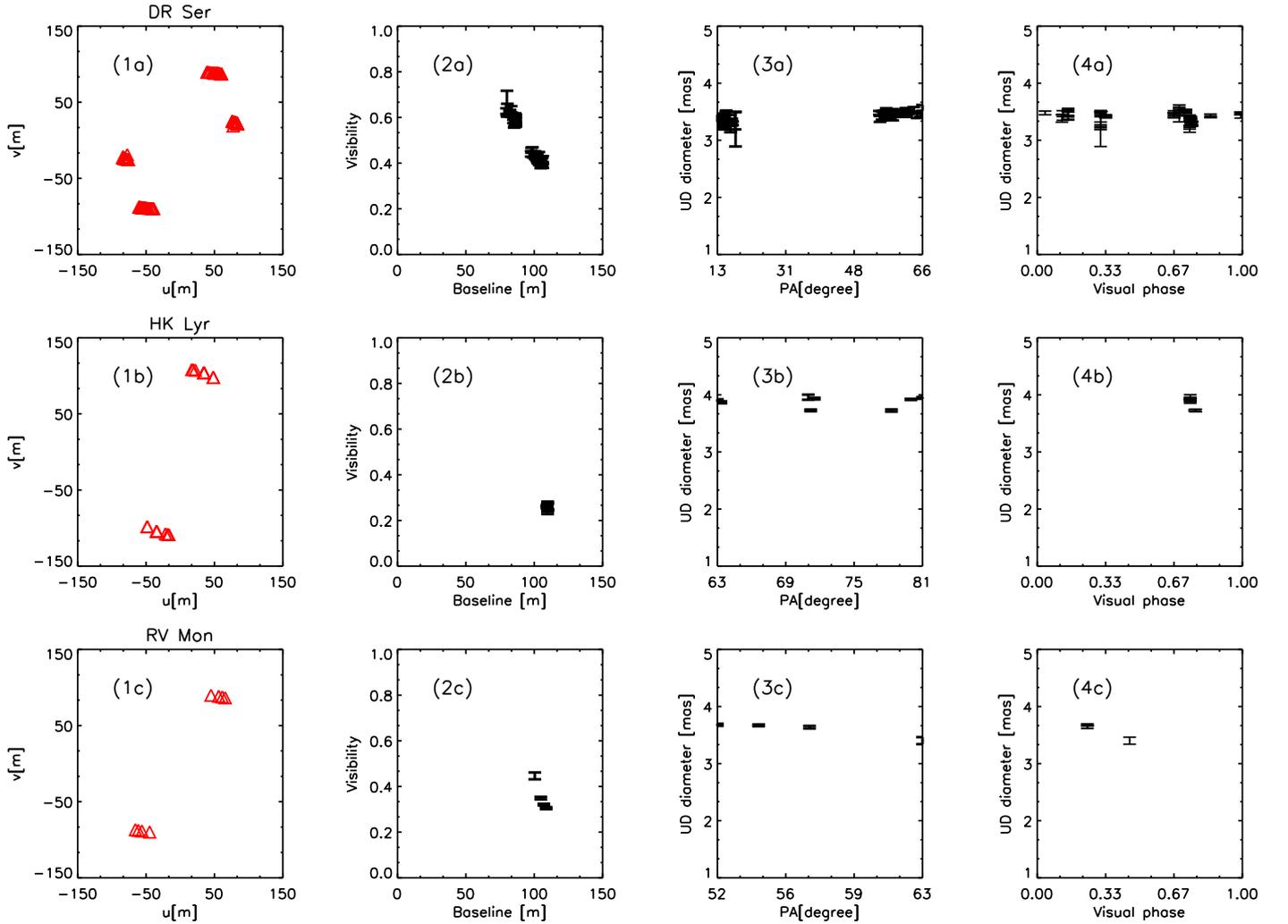}
           \caption{Checking of the hydrostatic and spherical symmetric approximation for the interferometric data.
           The series of panels 'a' is referred to DR~Ser data, panels 'b' to HK~Lyr, and panels 'c' to RV~Mon.
           The plots in the column '1' represents the $uv$-coverage of the observations, in column '2' 
           the visibility points acquired are plotted versus the baselines. 
           In column '3' the correspondent UD diameter 
           is plotted versus position angle.In column '4' the UD diameter is plotted versus the visual phase.}
           \label{ud.fig}
            \end{figure*}

From the relationship between visibility and uniform disk (UD) angular size,
$V^2 = [2 J_1 (x) / x]^2$, where $x = \pi B \theta_{UD} \lambda^{-1}$, we may
establish uniform disk angular sizes for the carbon stars observed by PTI since
the accompanying parameters (projected telescope-to-telescope separation, or
baseline, $B$ and wavelength of observation $\lambda$) are well-characterised
during the observation.  This uniform disk angular size will be connected
to a more physical limb darkened angular size in Sect.~\ref{pardet.sect}. 
   By plotting the UD diameter versus time, we could check the reliability of our approximation with hydrostatic models.
   From the UD diameter versus position angle the assumption of spherical symmetry is checked. 
   In Fig.~\ref{ud.fig} we present the interferometric observations for three of our five targets. The missing objects (CR~Gem, Z~Psc)
   have only one measurement available. The panels in the first row, identified with the letter 'a' represent the measurements for DR~Ser.
   In the row 'b' the points for HK~Lyr are plotted, and in the row 'c' there are the points for RV~Mon.
   In column '1' we present the $uv$ coverage. In column '2' the visibility points are presented versus the baseline. HK~Lyr and RV~Mon
   are observed always with the same baseline setup. The variation in the projected baseline is due to the Earth rotation.
   The UD diameter is plotted versus position angle in column '3'. 
   In column '4' the UD diameter are presented versus the visual phase in order to investigate eventual variations in the 
   angular size which might be due among other reason to the pulsation of the stars.
   The phase estimation for RV~Mon and DR~Ser is based on the visual light curve from the ASAS database \citep{poj02}; while 
   for HK~Lyr it is based on the light curve from the Hipparcos Variability Annex \citep{bei88}. 
   Small trends can be identified in the case of RV~Mon and HK~Lyr, they may be due to cycle-to-cycle variation, but 
   we can also not exclude completely an instrument issue (see detailed discussion in Sect.~\ref{rvmon.sect}).

   \section{Hydrostatic models and synthetic observables}
   \label{mod.sect}
   For the determination of the parameters we use the new grid of hydrostatic and spherically symmetric
   model atmospheres of \citet{ari09}. 
   They are computed with COMARCS, a program based on the MARCS code \citep{gus75, gus08}
   in the version used by \citet{jor92}, and \citet{ari97}.
   The models are generated assuming hydrostatic local thermal
   and chemical equilibrium. 
   The molecular and atomic opacity is treated in the opacity sampling (OS) approximation.
   The parameters that characterise a model are: effective temperature, metallicity, log$(g)$, mass, and C/O. 
   
   \subsection{Synthetic spectra}
   Our observed spectra have higher resolution than the ones of the grid published 
   in \citet{ari09}, therefore new synthetic spectra were computed for a subset of the initial 
   grid of hydrostatic models.
   
   We assumed solar metallicity for our targets. The subset of new spectra
   covers the following parameters \footnote{In the model grid the lower limit of log$(g)$ varies according to the temperature.
     For more details see Table~1 in \citet{ari09}.}: $2\,400\le T_{\rm{eff}}\le4\,000$ K with steps of $100$ K; $Z/Z_{\sun} = 1$;
   $-1.0\le $log$(g[$cm\,s$^{-2}]) \le +0.0$; 
       $M = 1 $\,or\,$2\, M_{\sun}$; C/O$ = 1.05, 1.10, 1.40$. 
   
   The COMA code \citep{ari00} was used to compute the opacities
   for the different layers of a given temperature-density atmospheric structure.
   The opacity sources included in the calculations for the continuum are listed in Table~1 of \citet{led09}.
   Voigt profiles were used for atomic lines, and Doppler profiles for the molecules.
   All the main molecular opacities typical for C-stars were included: CO \citep{gor94}, C$_2$ \citep{que74}, 
   HCN \citep{har06}, CN \citep{jor97} in the form of line lists, while C$_2$H$_2$ and C$_3$ \citep{jor89} as OS data.
   This is in agreement with the previous works \citep{loi01, ari09, led09} with the exception of the C$_2$
   linelist which was not scaled in the infrared range.
   
   The scaling of the $gf$ values for C$_2$ was suggested for the first time by \citet{jor97} and afterwards
   introduced by \citet{loi01} for fitting spectra of C-rich stars in the range of $0.5-2.5$ $\mu$m. 
   The authors keep the original linelist up to $1.15\,\mu$m, they scale it by a factor of 0.1 beyond
   $1.5\,\mu$m, and use a linear extrapolation in the transition region. 
   Different authors pointed out already the need of new C$_2$ line data.
   \citet{ari09} showed that the C$_2$ scaling does not affect so much the model structure, 
   while it introduces a variation in the spectral range between $1.3$ and $2.1\,\mu$m.
   In the same work a discrepancy was observed when comparing synthetic and observed 
   $(H-K)$ colours. This discrepancy was investigated in terms of C$_2$ opacity.
   The authors show in their Fig.~15 how the photometry obtained for unscaled models is in better agreement with observations.
   \citet{ari09} conclude that this could be an indication that the use of the linelist 
   in its original version is more appropriate, as long as there
   are no other more precise sources available. 
   
\begin{figure}[th]
     \centering
     \includegraphics[angle = 90,width = \hsize]{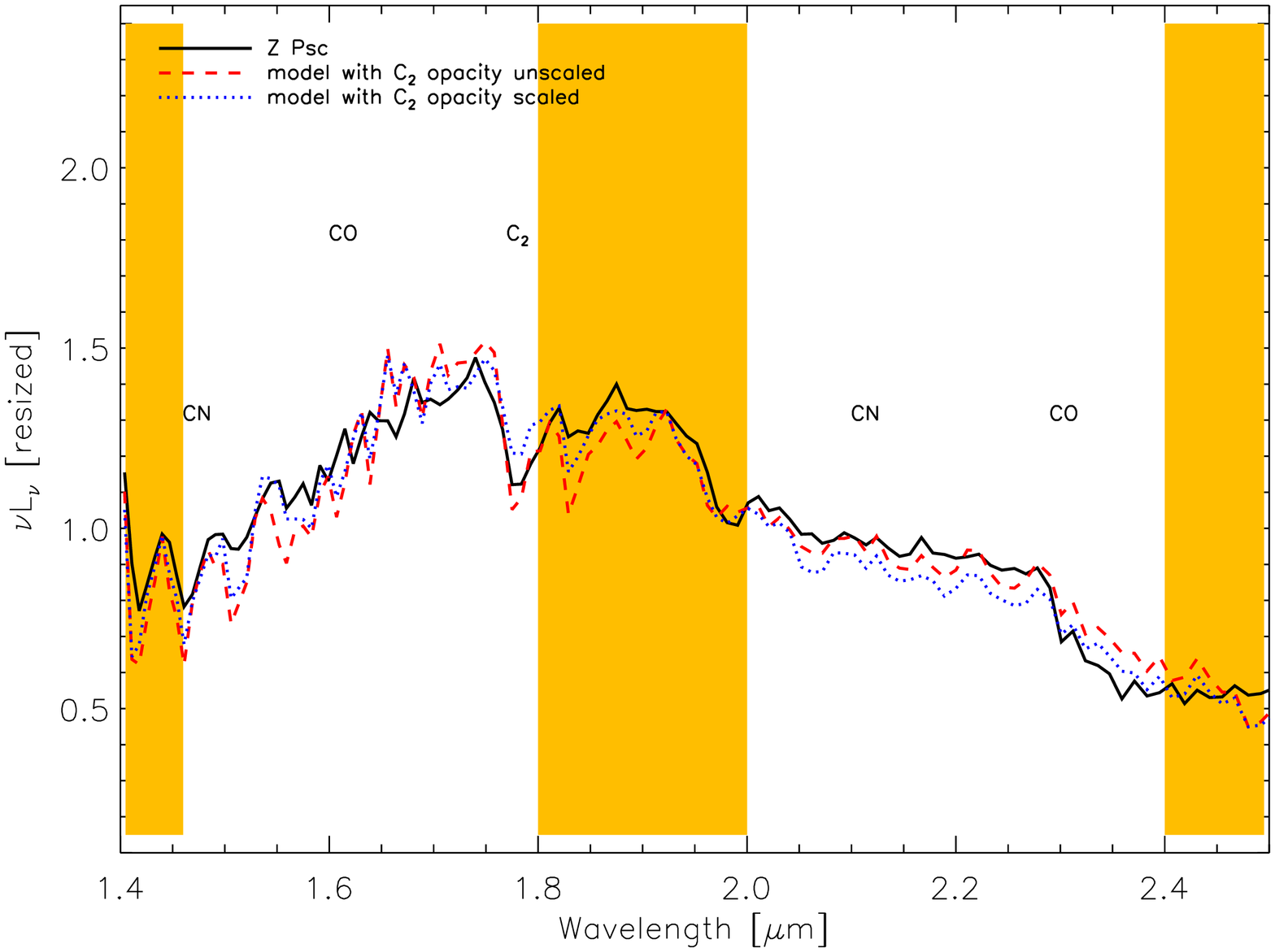}
     \includegraphics[angle = 90,width = \hsize]{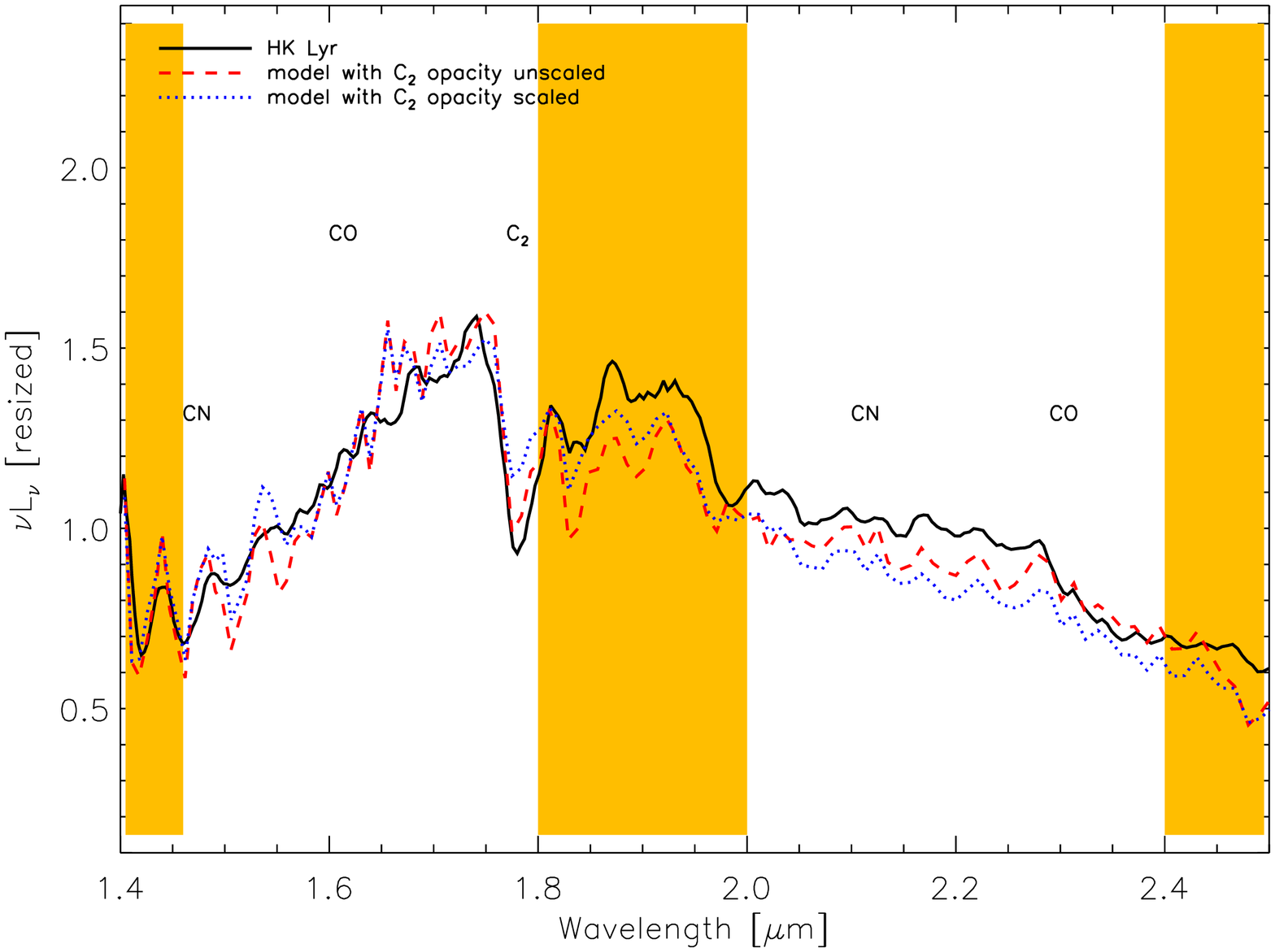}
     \caption{Comparison between the observed spectrum of two targets
       (black full line, Z~Psc in the upper panel and HK~Lyr in the lower), 
       a spectrum computed assuming scaled C$_2$ opacity (dotted line)
       and the spectrum with unscaled C$_2$ (dashed line).
     Shadowed bands mark regions of poor atmospheric transmission.}
     \label{c2op.fig}
   \end{figure}

   In order to check this conclusion further, we performed the $\chi^2$ test
   as explained in \cite{gar07}, and \cite{utt10}. In most of the cases we obtained a lower result of the
   $\chi^2$ test for the C$_2$-unscaled linelist of \citet{que74}. This result was also judged by eye
   in order to test the method.
   In Fig.~\ref{c2op.fig} the effect of the C$_2$ scaling in the $HK$ band is shown.
   The full line is the spectrum of the star Z~Psc (upper panel) and HK~Lyr (lower panel) convolved to a resolution of 200. 
   The dotted line is the synthetic spectrum 
   corresponding to a model with scaled C$_2$ opacity, while the dashed line is the same model with the original C$_2$ list.
   The result of the $\chi^2$ test, the check by eye, together with arguments of previous work of \cite{ari09} led us to opt for the
   original C$_2$-unscaled linelist.

   The resulting opacities are used as input for a spherically symmetric radiative transfer code
   which gives the synthetic spectra with a resolution of 18\,000 in the wavelength range $0.8-25\,\mu$m.
   We convolved these spectra to get same resolution as the observed data.
   
   \subsection{Synthetic visibility profiles}
   The spherical radiative transfer code produces an additional output beside the total spectrum: 
   monochromatic spatial intensity profile. 
   Synthetic visibility profiles for the $K$-broad band PTI setup 
   were computed for a subset of the grid of hydrostatic models.
   The metallicity, the temperature range and the C/O ratio were fixed (more details in Sect.~\ref{sect.TeffCO}).
   To compute the visibility profiles, i.e. the interferometric observables, we proceeded as follows.
   
   We defined a set of narrow-band filters centred on the sampled wavelength of the 
   transfer function for the PTI $K-$broad band setup. Then
   the monochromatic intensity profiles were convolved with the filters so defined.
   The visibility profiles were computed as Hankel transform of the
   narrow-band intensity profiles. The spatial frequencies were converted from AU$^{-1}$
   to baselines in meters, for a given distance.
   The squared visibility broad band profile finally is obtained as:
   \begin{equation}
     V_{\mathrm{broad}}^2 = 
     \frac{\sum\limits_{{i}}(S_{{i}}^2 F_{{i}}^2 V_{{i}}^2)}
          {\sum\limits_{{i}}(S_{{i}}^2 F_{{i}}^2 )}\,,
          \label{broad.formula}       
   \end{equation}
   were the sum was calculated over the $'i'$ narrow band filters; $S_i$ is the transfer function
   of PTI in the $K$ broad band setup; $F_i$ is the flux integrated over the narrow band filter, and
   finally $V_i$ is the visibility corresponding to the narrow band filters.
   Following this approach for the computation of the broad band visibility profile,
   the bandwidth smearing effect is properly taken in account \citep{ker03, ver05, pal09}.

   \section{Parameter determination}
   \label{pardet.sect}
   All the spectra of the targets appear very similar (Fig.~\ref{allspec.fig}). 
   They are dominated by the CN, C$_2$  and CO bands between 0.9 and $2.5\,\mu$m, while 
   the main absorption feature at $3\,\mu$m in the $L-$band is due to HCN and C$_2$H$_2$.
   In Fig.~\ref{allspec.fig} the regions affected by telluric absorption are shadowed.
   These regions were of course not considered in our fitting procedure.
   \citet{loi01} computed the flux ratio of various hydrostatic models (e.g. Fig.~3 of their paper)
   to investigate the effect of changes in the stellar parameters.
   They demonstrated that spectral features are mainly sensible to $T_{\rm{eff}}$ and C/O. 
   Moreover, \citet{ari09} showed that the synthetic spectrum below $2.5\,\mu$m of stars with $T_{\rm{eff}} \ge 3\,000$ K mildly
   changes for different values of log$(g)$ and mass.
   
   Therefore we determine $T_{\rm{eff}}$ and C/O from spectroscopy, and we turn to the PTI interferometric observations
   to determine the remaining characteristic parameters of the models: mass and log$(g)$.

   \subsection {Temperature and C/O ratio}
   \label{sect.TeffCO}
   
   The ratio between the $3\,\mu$m (C$_2$H$_2$) and the $5.1\,\mu$m (C$_3$) features is the most powerful tool 
   for determining temperature and C/O ratio for hydrostatic C-stars \citep{jor00}.
   Unfortunately our spectral coverage does not reach the C$_3$ feature at $5.1\,\mu$m.   
   
\begin{figure}
      \includegraphics[angle = 90,width = \hsize]{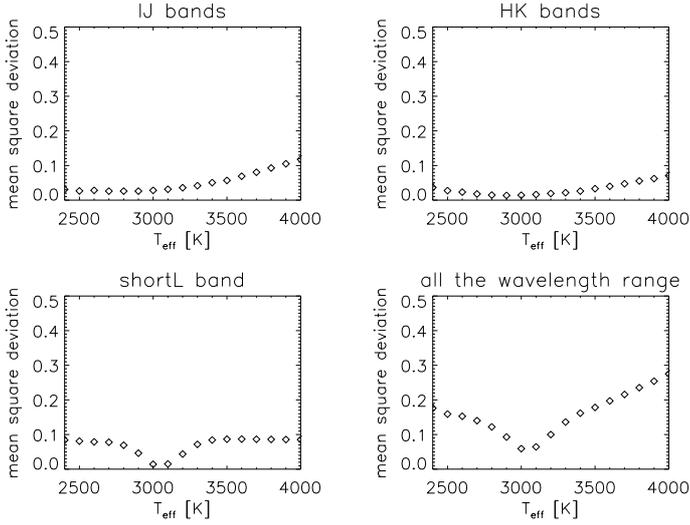}
      \caption{The four plots show the mean squared deviation obtained by fitting single portions of the spectra ($IJ$, $HK$,
        short$L$), and by fitting all the portions of spectra at once (all the wavelength range) for HK~Lyr.
        These plots demonstrate the accuracy of the short$L$ measurement.}
      \label{prob.fig}
    \end{figure}

   Following the approach of \citet{loi01} 
   we use the energy distribution of the available spectra as main temperature indicator.
   The observed spectrum was convolved to a lower resolution of $R = 200$.
   Every target of our list was compared with a grid of synthetic spectra with the same resolution.
   The grid of models is dense enough in temperature to allow an accurate statistic approach which is described in detail in 
   Appendix~\ref{temp.app}.
   The $T_{\rm{eff}}$ value was estimated for every wavelength range separately,  
   except for the long$L$ band. 
   This range of the spectrum is mostly flat and free from features therefore not very
   sensitive for the temperature estimation.
   We computed also a temperature taking into account all the
   four parts of the spectrum. 
   The resulting temperatures are presented in Table~\ref{teff.tab}. 
     Figure~\ref{prob.fig} shows the mean squared deviation\footnote{One root mean square value (rms) was derived by fitting each model. 
       The models were divided in bins of temperature; the single rms values were squared (mean square deviation) 
       in order to be summed for obtaining one value for every bin.
       The number of models in every bin is not constant, therefore the final mean square deviation was normalised 
       by the number of models in every bin.} derived by fitting first single pieces of the spectrum (``$IJ$'', ``$HK$'',
     ``short$L$'') and then by fitting all the pieces of the spectrum at once (``all the wavelength range'') for HK~Lyr.
      Our intention is to look for possible minima in the temperature distribution.

     The temperature values obtained from the different ranges of wavelength
     show a trend: by using the $IJH$ bands as indicators, cooler temperatures are obtained.
     This trend can be hardly explained with problems during the calibration. The data reduction procedure
     (see Sect.~\ref{spec.sect}) was accurate enough that we do not expect, in the overall flux distribution, systematic errors 
     which are large enough to explain the systematic variations in the temperature. 
     The temperature difference might be explained as the effect of an optically thin dust shell 
     surrounding the star. The effect of a dusty shell on the energy distribution of a star is shown in
     Fig.~3,4 of \cite{now11}. The model used for this simulation correspond to a very dynamic star. Our objects 
     are rather static, but we speculate that a thin dust shell would affect the spectrum in the same way with scaled-down intensity.
     This shell would absorb the light in the $IJH$ band 
     and emit it at longer wavelength. As a direct consequence a temperature determination based on $IJH$ band is systematically lower
     as can be seen in the upper panels of Fig.~\ref{prob.fig}.   
     If the shell is thin enough the effect on the short edge of the $L$ band is negligible.
     This makes the short$L$ band temperature determination quite robust, as can be seen in the lower left panel of Fig.~\ref{prob.fig}.

   \begin{table*}[!th]
       \caption{\label{teff.tab}Summary of temperature, C/O determination, and rms values corresponding to the
         different C/O.}
         \centering
       \begin{tabular}{l l l l l l l l}
         \hline
         \hline
         ID &  $T_{\rm{eff}} (IJ)$ & $T_{\rm{eff}} (HK)$ & $T_{\rm{eff}} ({\rm{short}}L)$ & $T_{\rm{eff}}$\,(all spec) &rms$_{1.05}$ & rms$_{1.10}$ & rms$_{1.40}$ \\
            & [K] & [K] & [K] & [K]&&&\\
         \hline                                                                                                                                                  
         CR~Gem & 2\,700 $\pm$ 200 & 2\,860 $\pm$ 200 & {\bf{3\,070 $\pm$ 50}} & 2\,920 $\pm$ 190  &                               {\bf 0.021} & 0.027 & 0.038 \\
         HK~Lyr & 2\,740 $\pm$ 220 & 2\,920 $\pm$ 210 & {\bf{3\,090 $\pm$ 50}} & 3\,080 $\pm$ 120  &                               0.015 & 0.016 & {\bf 0.013} \\
         RV~Mon & 2\,950 $\pm$ 300 & 2\,930 $\pm$ 230 & {\bf{3\,170 $\pm$ 50}} & 3\,210 $\pm$ 140  &                              {\bf 0.006} & 0.008 & 0.009 \\
         Z~Psc  & 3\,000 $\pm$ 330 & 3\,080 $\pm$ 210 & {\bf{3\,130 $\pm$ 60}} & 3\,170 $\pm$ 130  &                              {\bf 0.014} & 0.017 & 0.016 \\
         DR~Ser$_{no\,redd\,correct}$ & 2\,790 $\pm$ 250 & 2\,820 $\pm$ 200 & {\bf{3\,080 $\pm$ 40}} & 3\,030 $\pm$ 170 &{\bf 0.011} & 0.013 & 0.014 \\
         \hline 
       \end{tabular}
       \tablefoot{The bold face corresponds to the best values of temperature and C/O. More details in Sect.~\ref{sect.TeffCO}.}
   \end{table*}

   We note that this is not the case if the star is a strongly pulsating variable (such as Mira). In this case the profile of the $3\,\mu$m feature, 
   which is the result of the superposition of molecular opacity and dust continuous emission \citep{gau04}, it will be sensitive to the
   dynamic processes of the atmosphere. The respective temperature derived for every star by fitting the short$L$ band will be assumed hereafter.\\
   
   The second quantity which mainly modifies the appearance of a C-rich spectrum
   is the C/O ratio. Unfortunately the relatively low resolution of the spectra available limits the precision
   of this measurement. 
   The CO band at $2.29\,\mu$m, the C$_2$
   bands at $1.02$ and $1.20\,\mu$m, and the C$_2$H$_2 + $HCN at $3\,\mu$m were considered as indicators for the C/O ratio.
   Following to \citet{loi01} the C$_2$ features longward of $1.20\,\mu$m 
   were not used because of the uncertainty in the C$_2$ opacity. 
   Two reference wavelengths were chosen to indicate start and end of every selected band.
   In low resolution spectroscopy no continuum window is available for C-stars \citep{pal09}, therefore
   a \emph{pseudo-continuum} was
   obtained by linear interpolation of the points at the edge of the chosen band. 
   The selected portion of spectrum was normalised to this pseudocontinuum.
   Following this approach we derived the equivalent width for the observed spectrum and for all
   models with the temperature determined from the previous step.
   Grouping the models in sets according to the C/O (i.e. every ``C/O group'' includes models with same temperature, 
   but different log$(g)$ and mass), a root mean square value was obtained
   for every C/O by comparing the observed equivalent width and the synthetic one.

   In the cols.~6,~7,~8 of Table~\ref{teff.tab} the resulting rms for the three C/O values (1.05, 1.10, and 1.40) is given.
   The minimum rms is evidenced in bold.
   All the stars of the sample have a low value of C/O ratio which corresponds to 1.05, except 
   for HK~Lyr where the C/O = 1.4.
    
   \subsection{Mass, log$(g)$ and distance}
   
   As already pointed out by other authors it is hard to detect the effect of mass and log$(g)$ from 
   low resolution spectroscopy of C-rich stars.
   This is confirmed in our series of plots  in Fig.~\ref{crgem.fig}-\ref{drser.fig} where models with the same parameters
   except mass and log$(g)$ are compared with the spectra of the targets.
   The temperature and C/O ratio of the models were determined following the procedure
   previously described. 
   
   Due to these considerations we decided to treat the observables of interferometry
   as completely independent quantities.
   As interferometry aims to measure the radius of the target, the distance becomes an important parameter.
   Unfortunately the distance measurements available for these objects 
   are rather imprecise (typical error of the order of 20\%) and often in contradiction with each other with differences
   between measurement relative to the same objects which are larger than errors.   
   The problem we face is degenerate since we have to deal with radius, distance, mass, and all these quantities are related to each other.
   
   We handle the problem in the following way. Three distances obtained with different methods were chosen from the literature for
   every object (see Sect.~\ref{spec.sect} and Table~\ref{dist.tab}), and a set of synthetic visibility profiles was scaled to every distance.
   The set has a fixed temperature, and C/O ratio. 
   We obtained for every distance the combination of stellar parameters 
   ($M$, log$(g)$) from the best fitting profile. 
   
\begin{table*}[!th]
       \caption{\label{dist.tab}Distances measurements and stellar parameters of the model which best fits spectroscopic and interferometric measurements.}
           \centering
       \begin{tabular}{l l l l l l}
         \hline
         \hline
         Distance       & CR~Gem                  & HK~Lyr                & RV~Mon                & Z~Psc                  & DR~Ser \\
         \hline
          
         $d_{\rm{Berg}}$ & 920 pc                      & {\bf{730 pc}}                  & {\bf 670  pc}               & 465  pc                & 1295 pc\\
                       & $L = 11\,000$ $L_{\sun}$     &$L = 7\,186$ $L_{\sun}$   &$L = 6\,400$ $L_{\sun}$ &$L = 4\,534$ $L_{\sun}$ &$L = 18\,050$ $L_{\sun}$ \\
                       & $T_{\rm{eff}} = 3\,100$ K     &$T_{\rm{eff}} = 3\,100$ K &$T_{\rm{eff}} = 3\,200$ K&$T_{\rm{eff}} = 3\,100$ K&$T_{\rm{eff}} = 3\,100$ K\\
                       & C/O = 1.05                   &C/O = 1.4              & C/O = 1.05            & C/O = 1.05              & C/O = 1.05\\
                       & $M = 2 M_{\sun}$              & $M = 2 M_{\sun}$        &$M = 1 M_{\sun}$        &$M = 2 M_{\sun}$         &$M = 2 M_{\sun}$ \\
                       & log$(g) = -0.40$               & log$(g) = -0.20$         & log$(g) = -0.40$         & log$(g) = +0.0$ & log$(g) = -0.60$ \\
         \hline
         $d_{\rm{Clau}}$ &{\bf 780 pc}                  & {\bf 900 pc}                 & {\bf 1000 pc}               & {\bf 600 pc}            & {\bf 990 pc}            \\
                       & $L = 9\,025$ $L_{\sun}$ &$L = 11\,389$ $L_{\sun}$ &$L = 12\,932$ $L_{\sun}$ &$L = 7\,170$ $L_{\sun}$ &$L = 11\,390$ $L_{\sun}$ \\
                      & $T_{\rm{eff}} = 3\,100$ K &$T_{\rm{eff}} = 3\,100$ K&$T_{\rm{eff}} = 3\,200$K&$T_{\rm{eff}} = 3\,100$ K &$T_{\rm{eff}} = 3\,100$ K\\
                       & C/O = 1.05              &  C/O = 1.40            & C/O = 1.05        &    C/O = 1.05          &     C/O = 1.05\\     
                       & $M = 1 M_{\sun}$        & $M = 2 M_{\sun}$        &$M = 2 M_{\sun}$       &$M = 1 M_{\sun}$ &$M = 2 M_{\sun}$ \\
                       & log$(g) = -0.60$        & log$(g) = -0.40$         & log$(g) = -0.40$      &  log$(g) = -0.50$&   log$(g) = -0.40$\\
         \hline
        $d_{\rm{Hipp}}$ &  $323^{+357}_{-111}$ pc          & $1\,369^{+671}_{-NN}$ pc     & $450^{+369}_{-140}$ pc      & $323^{+119}_{-68}$ pc   & $1\,690^{+2810}_{-1120}$ \\
         & - & -                                                          &$L = 2\,574$ $L_{\sun}$ &$L = 2\,267$ $L_{\sun}$ & -  \\         
              &  -     & -                                        &$T_{\rm{eff}} = 3\,200$ K&$T_{\rm{eff}} = 3\,100$ K&-\\
                       &        &                                          & C/O = 1.05& C/O = 1.05 &\\
                       &        &                                          &$M = 1 M_{\sun}$         &$M = 1 M_{\sun}$ &\\
                       &        &                                          & log$(g) = +0.0$         & log$(g) = +0.0$ & \\
         \hline
       \end{tabular}
\tablefoot{The bold face distances correspond to the most likely values found by our stellar parameter analysis.}
   \end{table*}
\begin{figure}[!h]
  \centering
  \includegraphics[angle = 90,width = \hsize]{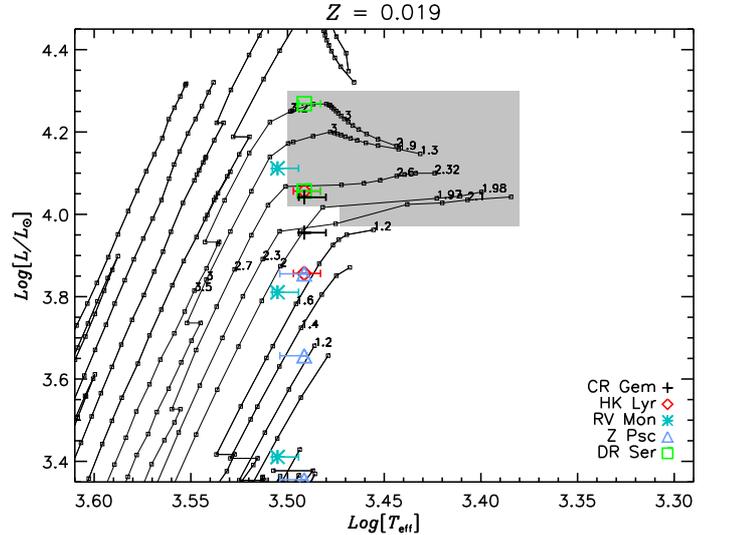}
  \caption{Zoom into the H-R Diagram where C-stars are located. 
    The solar metallicity isochrones \citep{mar08} are plotted in grey
    and the small numbers indicate the predicted present mass.
    The sampling of points in the isochrones is shown with tiny squares.
    The position of the C-stars is indicated by a shadowed area, the uncertainty about this area
    is discussed in Sect.~\ref{evol.sect}.
    The different symbols plotted correspond to the parameters
    determined for every star and distance assumed: plus for CR~Gem, diamond for HK~Lyr, star for RV~Mon,
    triangle for Z~Psc, and square for DR~Ser.}
  \label{track.fig}
\end{figure}
\section{Comparison with evolutionary tracks}
\label{evol.sect}

At this point of our investigation we have one combination of stellar parameters for every object and every distance.
To constrain the choice of the parameters further, 
we compare them with the recent isochrones for thermally pulsing AGB stars from \citet{mar08}.
The selection of the isochrones follows the same criteria as \citet{mar08}: ages log$(t/yr)$ between 7.8 and 10.2, 
and the spacing in log$(t)$ is 0.1 dex.

Fig.~\ref{track.fig} shows a zoom into the region of the H-R diagram where AGB stars are located.
The shadowed area identify the region where C-stars are expected.
Due to the discrete sampling of the points in the isochrones (small square), the transition region 
from M- to C-stars is not very well defined. According to the stellar evolution calculations (P. Marigo, private communications)
all the stars on the cool side of the ``hook'' of the isochrones are C-stars.
Some values of the present mass are marked on the isochrones: they will be compared with the mass 
of the best fitting hydrostatic models (see Sect.~\ref{disc.sect}). 

We overplot to the isochrones the points corresponding to the parameters we determined. 
Every star is indicated with a different symbol centred on the temperature and luminosity
associated to the model: a plus for CR~Gem, an asterisk for HK~Lyr, diamond for RV~Mon, triangle for Z~Psc, and square for DR~Ser. 
The error bar of the temperature 
are the ones associated to the short$L$ band $T_{\rm{eff}}$ determination, therefore they are centred on the $T_{\rm{eff}}({\rm{short}}L)$ 
values given in col.~3 of Table~\ref{teff.tab}.
The errors on the luminosity  are not plotted here to avoid confusion but they are of the order of $40\%$ (corresponding to the
uncertainty on the distance measurements $d_{\rm{Berg}}$, $d_{\rm{Clau}}$).

\section{Discussion on individual targets}
\label{obj.sect}
\subsection{CR~Gem}
CR~Gem is classified as an irregular Lb variable with a variability amplitude in the $B$ band of 1.20 mag in the General Catalogue
of Variable stars \citep[GCVS;][]{sam09}.
Nevertheless \citet{whi08} classify this star as SRb with a period of 250$^d$ and the ASAS $V$ band light curve shows an amplitude of 0.5 mag \citep{poj02}.
The spectral classification given in the GCVS is C8,3e(N). 

\begin{figure*}
  \centering
  \includegraphics[angle = 90,width = \hsize]{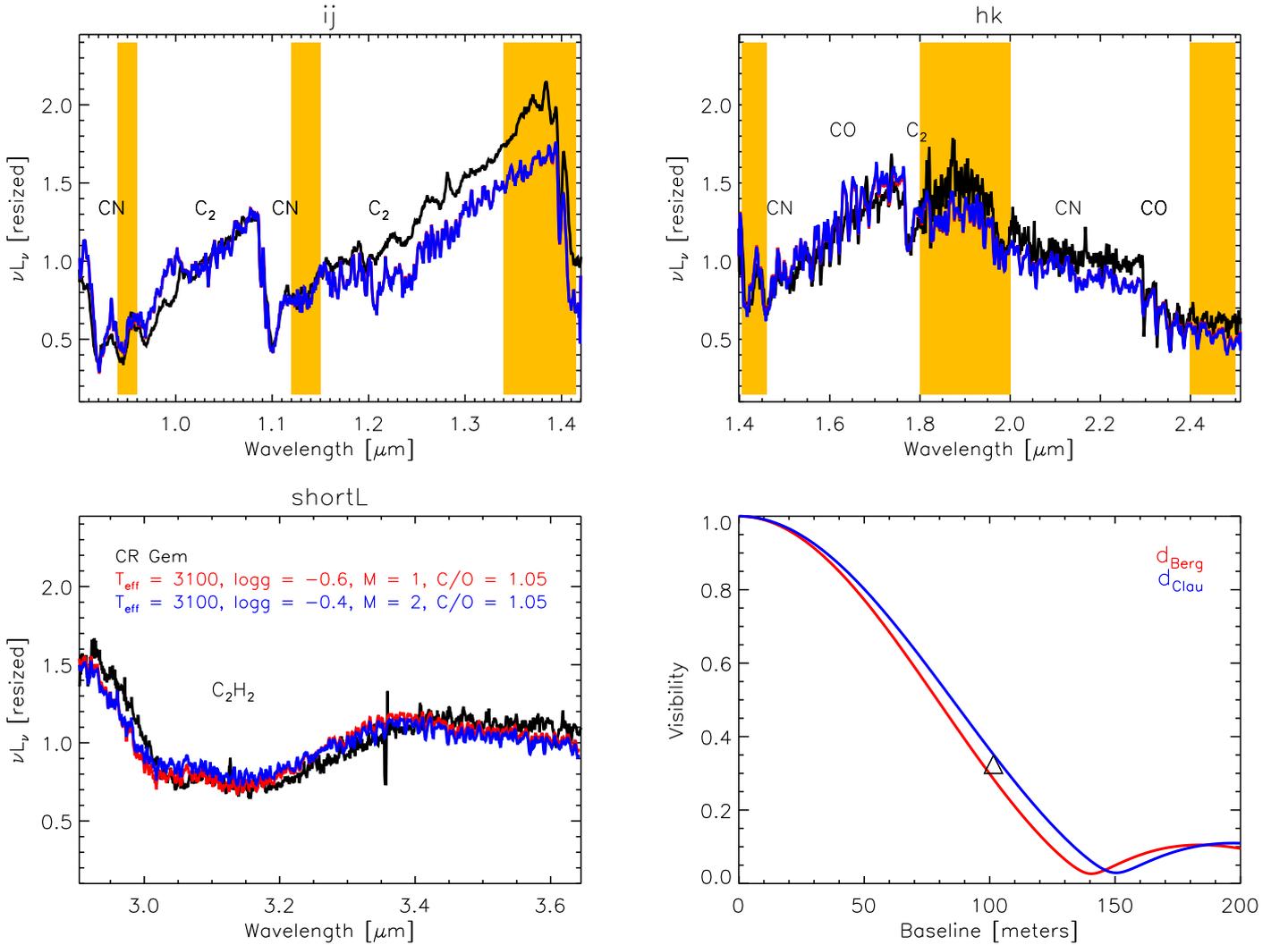}
  \caption{Comparison of the UKIRT/UIST spectra (black line) and PTI interferometric measurements (black triangle) of CR~Gem with
    hydrostatic models predictions. The upper left panel illustrate the $IJ$ range of the spectrum, the upper right panel shows the
    $HK$ wavelength range. The lower left panel shows the short$L$ range. The shadowed bands mark the region with poor atmospheric transmission.
    The molecules that contribute to the spectrum are also indicated. The synthetic spectra are overplotted in grey (blue, and red in colour version).
    The lower right panel shows the interferometric data point with the models which best fit the data for different distances overplotted.}
  \label{crgem.fig}
\end{figure*}

The $3\,\mu$m short$L$ temperature we determine is around $100\,\rm{K}$ higher than the $2\,960\,\rm{K}$ measured by \citet{ber05}. . 
The C/O we get is 1.05, we could not find other estimates for this value in literature.
The observed spectrum is presented in Fig.~\ref{crgem.fig}. It is 
in general well reproduced with a few exceptions:
the range $1.2-1.4\,\mu$m and the $K$-band between $2.1$ and $2.3\,\mu$m where the 
real data show an excess compared to the models (see general discussion in Sect.~\ref{disc.sect}).

For this star we collected only one point of PTI visibility (Fig.~\ref{crgem.fig} lower right panel), therefore it is not possible 
to check for asymmetries.
All the visibility profiles corresponding to the Hipparcos distance are not extended enough to fit the visibility point, 
therefore this distance can be excluded.
For the Claussen distance the best fitting model has mass $1\,M_{\sun}$ and log$(g) = -0.60$, the luminosity is $9\,000\,L_{\sun}$ 
and the radius of the star $330\,R_{\sun}$.
For the Bergeat distance, 
the largest value of distance estimated for this star, the best fitting model has $2\,M_{\sun}$ and log$(g) = -0.4$,
the luminosity is $11\,000\,L_{\sun}$ and the radius $370\,R_{\sun}$.

\subsection{HK~Lyr}
According to the GCVS the visual amplitude of HK~Lyr is 1.80 mag, the spectral classification C6,4(N4),  
and the star is classified as irregular Lb. The Hipparcos Variability Annex \citep{bei88} reports for this star 
a period of 196$^d$ and an amplitude of variability of 0.4 in the Hipparcos magnitude.
From optical spectroscopic analysis this star is enriched in Lithium and Technetium \citep{abi02, bof93}.

\begin{figure*}
  \centering
  \includegraphics[angle = 90,width = \hsize]{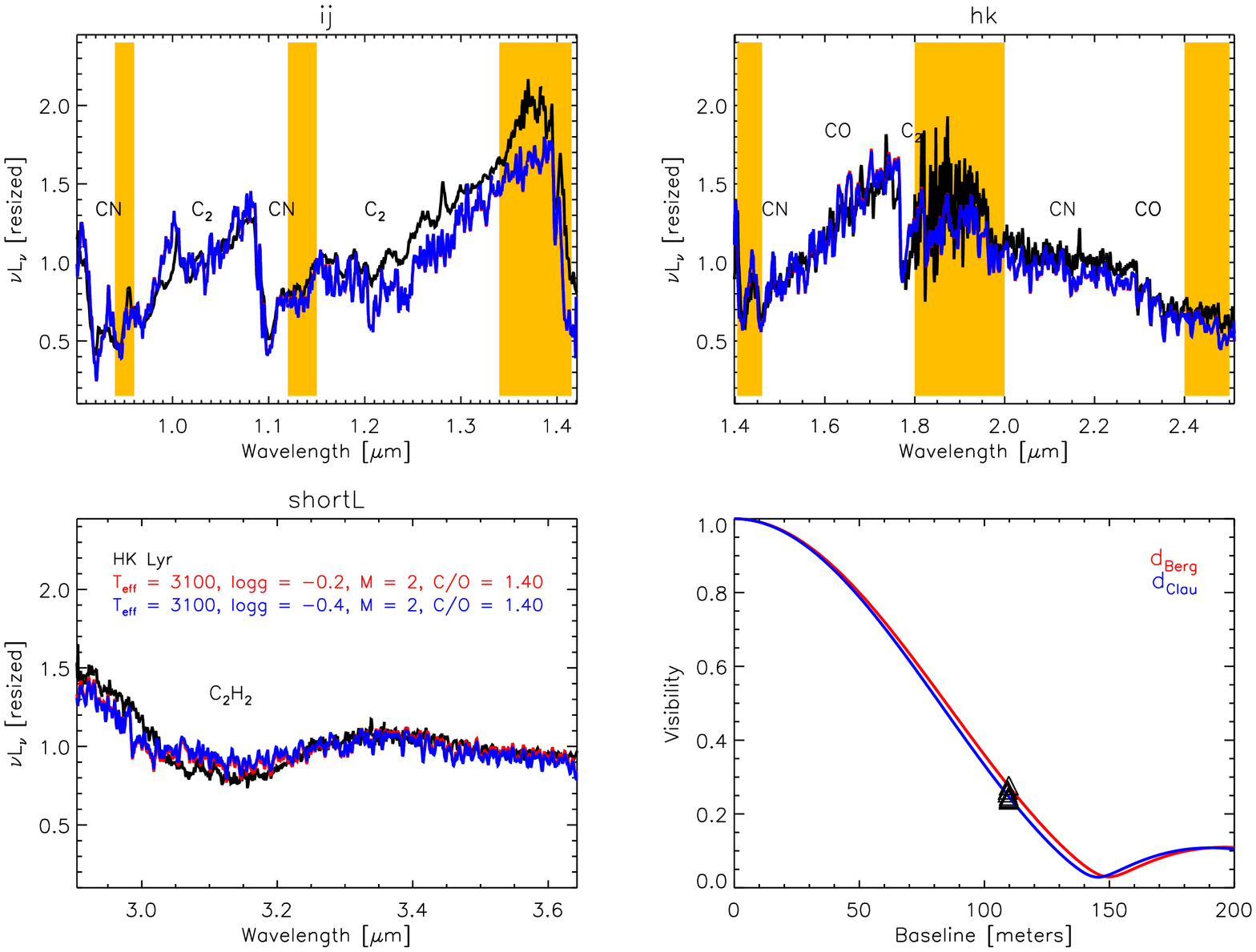}
  \caption{Same as in Fig.~\ref{crgem.fig} for HK~Lyr.}
  \label{hklyr.fig}
\end{figure*}

From our spectroscopic investigation the temperature and C/O ratio obtained are higher 
than the ones estimated in literature (Table~\ref{teff.tab}).
While our temperature is $3\,080\,{\rm{K}}$, \citet{ohn96} give $2\,866\,{\rm{K}}$ and \citet{ber05} estimate $2\,945\,{\rm{K}}$. 
Concerning the C/O ratio we obtain 1.4 while \citet{abi02} estimate 1.02.
The fit of the spectroscopic data is quite successful 
except in the region between 1.2 and 1.45 $\mu$m (Fig.~\ref{hklyr.fig} upper left panel).

Eight visibility points are available from PTI observations.
The UD diameter computed by fitting the single points do not differ from each other.
Over the time interval and position angle of the observations, the star does not show notable variation nor evidence of asymmetries
(Fig.~\ref{ud.fig}, row 'b').
We fitted all the visibility points with synthetic profiles scaled for 
different distances (Fig.\ref{hklyr.fig} lower right panel) . 

None of the models corresponding to the Hipparcos distance can fit the points, therefore we excluded this distance. 
The best fitting parameters corresponding to $d_{\rm{Berg}}$ are $2\,M_{\sun}$ and log$(g) = -0.2$.
This model has a luminosity of $L = 7\,200$ $L_{\sun}$ and a radius $R = 300$ $R_{\sun}$.
The best fitting model corresponding to $d_{\rm{Clau}}$ has again mass 2 $M_{\sun}$ but log$(g) = -0.4$.
This model has a luminosity of $L = 11\,400 L_{\sun}$ and a radius $R = 370$ $R_{\sun}$.
In Fig.~\ref{track.fig} the resulting parameters  for HK~Lyr are plotted with asterisks.

\subsection{RV~Mon}
\label{rvmon.sect}
RV~Mon is a C4,4-C6,2(NB/R9) 
star, its variability class is SRb with a primary pulsation period of 131$^d$ and a long secondary period
detected by \citet{hou63} of 1047$^d$. 
The amplitude of variability given in the GCVS is 2.19 mag in the $B$-band, while the ASAS light curve 
suggests a $V$ amplitude of 0.3 mag.

The temperature obtained with our fitting procedure is in between the values given in the literature. 
\citet{ohn96} get $T_{\rm{eff}} = 3\,330\,{\rm{K}}$, we obtained $3\,170\,{\rm{K}}$ while \citet{ber05} give $2\,910\,{\rm{K}}$.
No estimation of C/O was found in the literature to be compared with our 1.05.
The model we adopted for RV~Mon matches quite well the spectroscopic observations 
in all the ranges except in the region of the CO bands (longward $2.29\,\mu$m) where
the observations show an IR excess compared to the synthetic spectra.
The intensity of the C$_2$ feature at $1.77\,\mu$m is not well reproduced but this is also at the border of an atmospheric window.

\begin{figure*}
  \centering
  \includegraphics[angle = 90,width =\hsize]{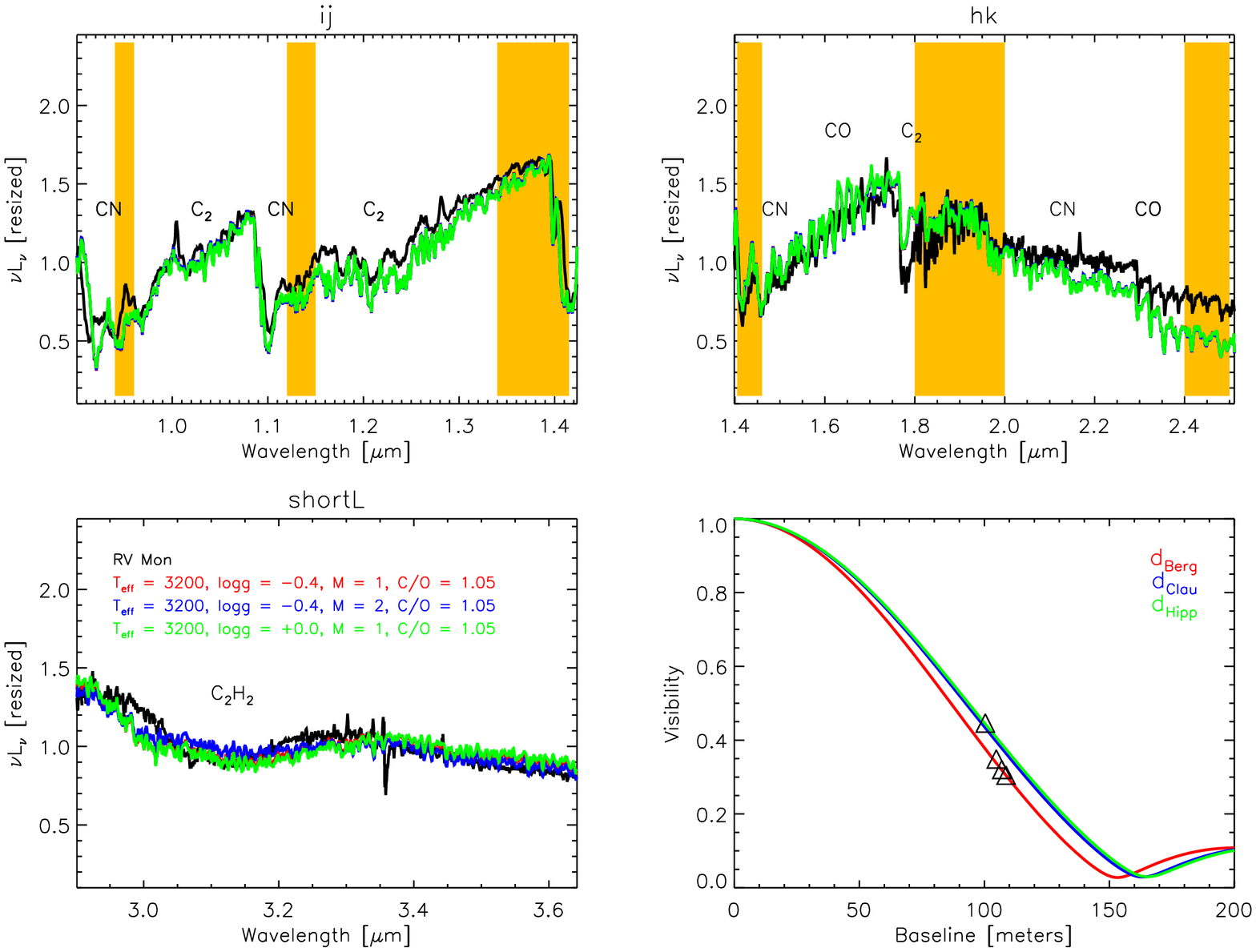}
 \caption{Same as in Fig.~\ref{crgem.fig} for RV~Mon.}
  \label{rvmon.fig}
\end{figure*}

For this star we collected 4 visibility points. The first one is in 1999, the other three in 2008.
Although the observations were carried out at very similar baseline, 
the visibility jumps by 0.1 between the two epochs of observation. 
This jump is visible also in the UD-diameter where it amounts to $\sim~0.16$ mas which 
is larger than the estimated error on the UD diameters
($\sim 0.06$ in the worst case). This difference is also noted in the plot of the UD versus position angle.
A careful check was performed to exclude any problem coming from the calibration procedure, but still 
we cannot exclude completely that the fluctuation in the visibility is due to an instrument issue.
From the astrophysical point of view the observed trend could be explained in different ways: 
(i) it could be an asymmetry plus a temporal variation (for example a large 
convective shell), (ii) or an effect of the pulsation (panel '4c'). 
The temporal variation due to the primary short period of variability is not enough to explain the jump in the visibility.
We derived from the light curve a difference in phase for the primary period of 0.16 between the two sets of observations. 
According to the predictions of the dynamic model atmospheres
\citep[see Fig.~7 and Fig.~9,]{pal09} a hot C-star should show a difference in UD-diam of 0.06 mas, definitely smaller than 0.16. 
The two sets of observations have been obtained with a time difference of 3.5 times the length of the long secondary period. 
This long term variation may, thus, be responsible for the observed variation in size. Further observations are necessary to reach a conclusive interpretation.
Keeping this in mind we performed a fit of the interferometric points with a hydrostatic model atmosphere
as a first step. This might be followed by further investigations.
From the lower right panel of Fig.~\ref{rvmon.fig} it is immediately obvious 
that no synthetic visibility profile can fit at the same time all the observations.
The best fitting model obtained for the Hipparcos or Claussen distances, can fit only the single point
observed in 1999. The other models are either too extended or not extended enough to fit the other observations. 
The best fitting model for 
$d_{Hipp}$ has $M = 1 M_{\sun}$, log$(g) = +0.0$, $L = 2\,600$ $L_{\sun}$ and $R = 170$ $R_{\sun}$.
For the distance determined by \citet{cla87} the single data point is fitted by the model with  
$2 M_{\sun}$, log$(g) = +0.4$, $L = 13\,000$ $L_{\sun}$ and a radius $R = 370$ $R_{\sun}$.
On the other hand for the distance determined by \citet{ber05} the three points observed in 2008 are fitted 
by the model with $M = 1 M_{\sun}$, log$(g) = -0.4$,
$L = 6\,400\,L_{\sun}$, and $R = 260$ $R_{\sun}$.
Nevertheless the effect of variability or possible asymmetries need to be investigated in more detail,
and the parameters derived for this object should be considered with caution.

\subsection{Z~Psc}
Z~Psc is an SRb variable with amplitude derived from photographic plates of 1.3 mag. The period in the GCVS is 144$^d$.
The ASAS light curve has an amplitude of 0.52mag in $V$ band.
The spectral classification from the GCVS \citep{sam09} is C7,2(N0). 
 \citet{abi02}, and \citet{bof93} determined a \element[][12]{C}$/$\element[][13]{C}$ = 55$, moreover they found enhancement of Tc, and no trace of Li
in the spectrum. This implies that Z~Psc is a standard low-mass TP-AGB star, and an intrinsic C-rich star. 

\begin{figure*}
  \centering
  \includegraphics[angle = 90,width = \hsize]{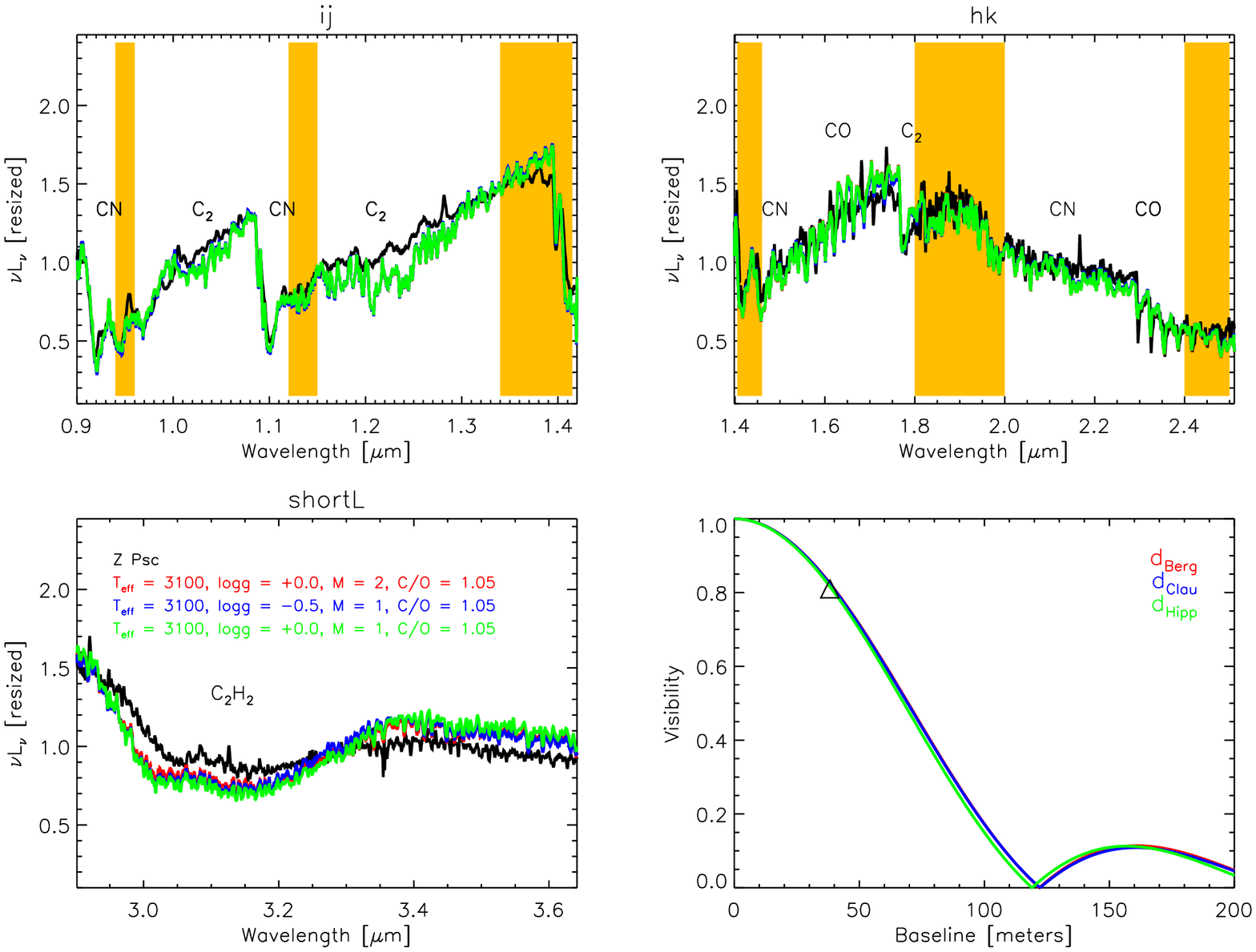}
  \caption{Same as in Fig.~\ref{crgem.fig} for Z~Psc.}
  \label{zpsc.fig}
\end{figure*}

Many temperature estimates were given for this star in literature:
\citet{lam86} obtained 2870 K, \citet{dyc96} 3240 K, \citet{ohn96} 3150 K, \citet{ber05} 3095 K.  
All these values were obtained with different methods.
Our measurement (see Table~\ref{teff.tab}) is very close to the one given by \citet{ohn96}.
The C/O = 1.05 we estimated is in agreement with the 1.014 of \citet{lam86}, and with the more recent 1.01 value of \citet{abi10}. 
The match between models and observations is shown in Fig.~\ref{zpsc.fig}.
From spectroscopy a small disagreement is barely detectable in the C$_2$ band at $1.20\,\mu$m, while the $3\,\mu$m feature of the model
is too deep to fit the data (the star is hotter than the model). 

Z~Psc is the only one among our targets with no PTI data.
For this star we used an observation available from the literature \citep{dyc96} taken in the $K$-broad band with
IOTA (2.2 $\mu$m) as already described in Sect.~\ref{interf.sect}.
The fit of the interferometric IOTA data is presented in the lower right panel of Fig.~\ref{zpsc.fig}.
The best fitting model for the Hipparcos distance has 
a mass $M = 1 M_{\sun}$, log$(g)$ = +0.0, $L = 2\,300\,L_{\sun}$, and $R = 165\,R_{\sun}$.
The distance determined by \citet{ber05} gives a best fitting model with the same log$(g)$ obtained for Hipparcos distance fit but
$M = 2\,M_{\sun}$, $L = 4\,500\,L_{\sun}$ and a radius $R = 230\,R_{\sun}$.
In the case of the Claussen distance we have the best fitting which is the model with mass $1\,M_{\sun}$, log$(g) = -0.5$, $L = 7\,150\,L_{\sun}$, and $R = 300\,R_{\sun}$.


\subsection{DR~Ser}
\label{drser.sect}
According to the GCVS, DR~Ser is an irregular variable Lb with an amplitude in the $B$-band of 2.99 mag.
The ASAS light curve of this object is quite stable over the last 2000 days, it has a period of 196$^d$, and a $V$ band amplitude of 0.3 mag.
The spectral class is C6,4(N).


This object is the one in our sample with the highest value of interstellar reddening.
Applying the reddening correction we noted that the observed spectrum looks completely unrealistic.
Remembering that reddening correction for single objects is always very uncertain
we decided to skip the step of the reddening correction for DR~Ser.
Further investigation and a detailed map of the ISM in the region of this star are needed.

We determined a temperature of $3080\,{\rm{K}}$. In the literature the $T_{\rm{eff}}$ values associated to this object are usually lower 
(see Sect.~\ref{disc.sect}):
$2500\,{\rm{K}}$ \citep{abi97}, $2570\,{\rm{K}}$ \citep{tho02}, $2650\,{\rm{K}}$ \citep{sch05, ber05}.  
\citet{abi97} estimated the isotopic ratio $\element[][12]{C}$/$\element[][13]{C} = 6$, which is also surprisingly low!
The C/O we found, 1.05 is lower than the C/O = 1.26 obtained in literature by \citet{egl95}.

\begin{figure*}
  \centering
  \includegraphics[angle = 90,width = \hsize]{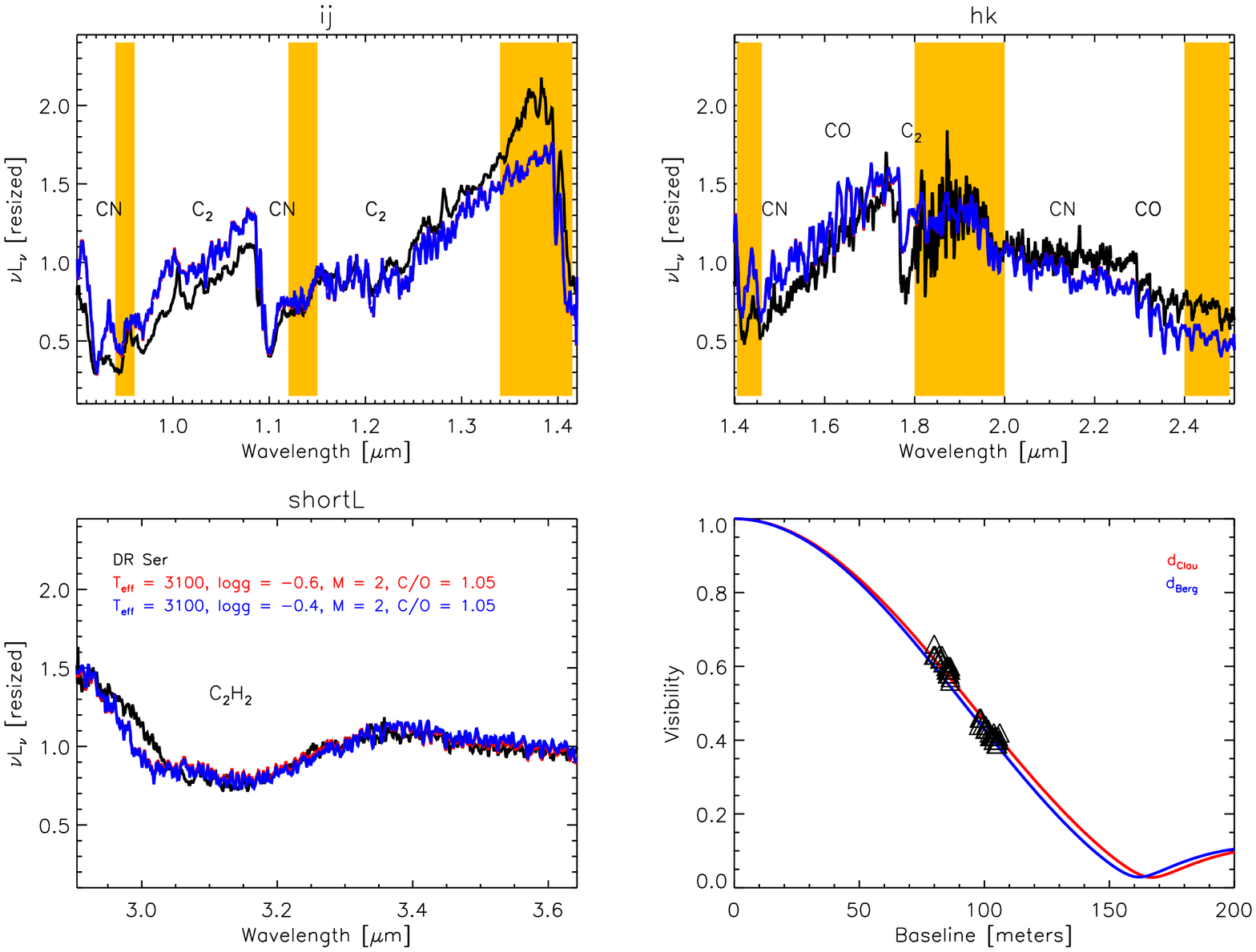}
  \caption{Same as in Fig.~\ref{crgem.fig} for DR~Ser.}
  \label{drser.fig}
\end{figure*}

There are 54 PTI points available for this object in a period between June 2000 and August 2001.
The UD diameter determined agree with each other within a range of $\pm 0.02$ mas, therefore we considered for the
model fitting all the points together (Fig.~\ref{drser.fig}).
None of the profiles computed at the Hipparcos distance could fit the data points.
The best fitting models 
have a mass $M = 2\,M_{\sun}$, and log$(g) = -0.4$ or $-0.6$ for $d_{Clau}$ and $d_{Berg}$ respectively.
The first model has $L = 11\,400\,L_{\sun}$ and a radius of $370\,R_{\sun}$; the second one 
$L = 18\,000\,L_{\sun}$ and $R =  470\,R_{\sun}$.
It is impressive to note how all the visibility points observed at different epochs and position angles
fits the hydrostatic synthetic profile.

\section{Discussion}
\label{disc.sect}
Based on the fitting procedure previously described,
we can identify a sequence in effective temperature for our targets.
The temperature increases starting from CR~Gem, DR~Ser, HK~Lyr, Z~Psc, RV~Mon.
Our temperature determination is very precise thanks to the large grid of models available 
and to the use of the short$L$ band spectrum.
The $T_{\rm{eff}}$ here determined are always compatible within the error given in literature.

The C/O ratio determination is not as accurate as the one of the temperature because of the very
coarse spacing in the model grid, but mainly because of the low resolution of the spectra. 
Although the C/O values we determine have to be considered as indicative,
they are quite in agreement with the ones estimated in literature.
We obtained the same C/O$ = 1.05$  for all the stars except HK~Lyr which has C/O$ = 1.4$.
Those values are in agreement with the range [1.01, 1.4] reported in literature for classical C-stars (see Fig.~42 of \cite{lam86}, and Table~2 of \cite{ber05}).

In general there is a good agreement between synthetic and observed spectra.
Small discrepancies are observed mainly in the region between 1.2 and 1.4 $\mu$m,
between 2.1 and 2.3 $\mu$m and on the left edge of the 3 $\mu$m feature.
They can be attributed to small calibration problems, but
also to a small amount of dust. This last point is supported by the fact that 
the differences detected in the $IJ$ band between 1.2 and 1.4 $\mu$m are larger for the lower temperature
stars (CR~Gem, HK~Lyr, DR~Ser). 
We simulated the effect on the spectrum of a small amount of dust by using the DUSTY code \citep{ive97}.
Indeed the $IJ$ band of the spectrum fits slightly better for stars which have a large $J-K$ value (CR~Gem, DR~Ser).
We also checked the effect of this small amount of dust on the stellar parameter determination.
The effect on the temperature and on the interferometric observable was within the error bar we already took into account, therefore we
proceed with the hydrostatic models.
A detailed analysis of spectro-interferometric observations with hydrostatic+dusty model atmosphere will be presented in a follow-up work.

From the fit of the interferometric observations with models at different distances we note
that in three over five cases no model can fit the observations at the distance estimated from Hipparcos.
This is not such a surprise considering the error associated to these distance measurements.
Further consideration about the distance can be made by studying the position of the best fitting models
in the H-R diagram (Fig~\ref{track.fig}).

Some of our hydrostatic models lie in a region of the H-R where higher masses are expected.
If we consider the agreement between the two masses (from hydrostatic models and evolutionary calculations) 
as a criterion to select the correct distance of the objects 
then we have the following:

\begin{itemize}
\item CR~Gem. The point with lower luminosity is the one fitting the range of mass, therefore the distance
determined by Claussen can be retained, and the one of Bergeat discarded.
\item HK~Lyr. Both the models have the same mass and lie in a region in between 1 and 2.6\,$M_{\sun}$.
 Considering an error on both the distances of $20\%$ on the distances (i.e. a factor 1.44 in luminosity) 
 none of the values can be excluded.
\item RV~Mon. The Hipparcos distance is in the correct region of mass, but too far from the C-stars region, therefore can be discarded.
Both the other two distances correspond to a model which fits the region of masses, therefore they can be both retained. 
\item Z~Psc. In this case the Bergeat distance can be discarded since the best fitting model has 2 $M_{\sun}$ , but falls in the 
range of 1 $M_{\sun}$. Considering the error on the luminosity both the distances from Hipparcos and
Claussen can be retained. Nevertheless the position of the model at the Claussen distance is more probable
since this is closer to the C-stars region.
\item DR~Ser. The distance estimated from Bergeat can be discarded since the model lies in the region of the stars with 3 $M_{\sun}$.
\end{itemize}

\begin{figure*}
  \centering
  \includegraphics[angle = 90,width = \hsize]{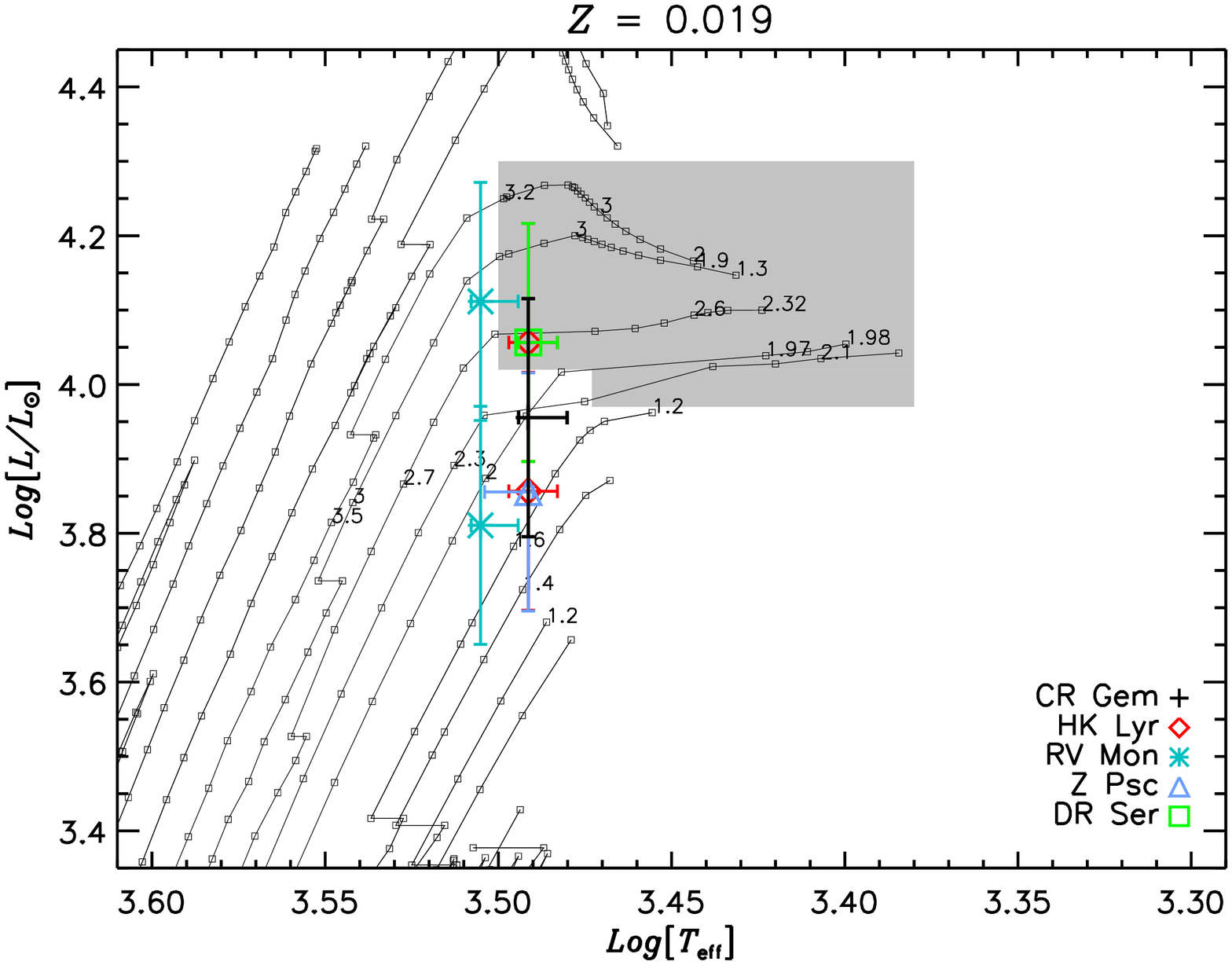}
  \caption{Same as in Fig.~\ref{track.fig}, but this time only the parameters with most probable distances are plotted.}
  \label{track-clean.fig}
\end{figure*}

The parameters with the most probable distances, with associated errors in temperature and luminosity are shown in Fig.~\ref{track-clean.fig}. 

Taking into account the uncertainty associated to the transition phase M-C stars (see Sect.~\ref{evol.sect}),
all the stars have at least one combination of parameters that
lie in the region where the C-stars sequence starts.
This is in agreement with the hydrostatic scenario
for our targets: the objects are supposed to be hot C-stars, with relatively low mass-loss and
low C/O. They ``recently'' turned into C-stars and while moving on the evolutionary track 
they will increase the production of carbon and will lose more and more mass because of the stellar wind.
Our work is partially limited by the fact that only model atmospheres with mass 1 and 2 are available.

Two objects of our sample show some peculiarity which deserve further investigations.
The interferometric observations of RV~Mon show a possible non-hydrostatic nature.
All the visibility points acquired with PTI cannot be fitted with the same model.
This might be due to some instrumental effect, while the physical explanation would be an evidence 
of an asymmetry or (more probably) of a variability effect.
According to the interferometric measurements, the star DR~Ser is an hydrostatic object.
We were able to fit with the same model 54 visibility points observed in different epochs
and position angles.
On the other hand the fit of the $IJ$ and $HK$ band is not very good,
and the reddening correction of the spectrum makes the fit even worse.
A detailed ISM map for this region of the sky is needed.
 
\section{Conclusions}
\label{concl.sect}
More than a simple parameters determination, 
this work aims to identify which of the stellar parameters of C-stars can be constrained
by using different techniques like spectroscopy and interferometry. 
Which limits appear by investigating only with one tool?
Infrared interferometry is a relatively young field of research, but it can already give a 
significant contribution on the stellar parameters determinations.
\\

In this work we confirm that low resolution infrared spectroscopy is not sensitive
to log$(g)$ and mass determinations of C-stars. 
We show clearly that a highly precise determination of the temperature 
can be achieved for C-stars with no significant dust contribution
thanks to the use of the $3\,\mu$m feature (C$_2$H$_2$ + HCN).
The determinations based only on the $IJH$ bands tend to underestimate the temperature.

The hydrostatic model atmospheres adopted in this work were able for the first time to fit 
simultaneously spectroscopic and interferometric observations.
This is a very important achievement considering that inaccuracies in model atmosphere
reflect negatively in the check of the stellar evolution calculations.
On the other hand the inaccuracy in the distance determination
for this class of objects reflects negatively on the parameter determination, as already noted in \cite{wit01}.
We considered a criterion to select the distance on the agreement between mass of the hydrostatic models 
and mass predicted by the isochrones.
According to this, the distance estimated by \cite{cla87} is the most probable among the three distances we used
for this work.
Taking into account also the uncertainty in the transition phase from O-rich to C-rich star in the evolution models,
we obtained that all the objects of our sample have at least one combination of observationally determined parameters which fits
the C-stars region.
We were able to associate temperature, C/O,
mass, log$(g)$ and a range of reasonable distances for all the objects of our sample.

A grid of models with larger coverage of the mass range (at least from 1 to 4\,$M_{\sun}$),
as well as a more dense grid of points for the isochrones,
would improve significantly our estimations.
The advent of a new dedicated mission (Gaia) might improve the distance problem for C-stars.

The synergy between different techniques of investigation and 
realistic theoretical atmospheric and stellar evolutionary models
will be the key to determine precise characteristics for
this class of objects, and to be able to understand clearly the physical processes
which drives their evolution.

\begin{acknowledgements}
  The authors acknowledge the anonymous referee for useful comments that helped improve the paper.
  This work is supported by the Projects P19503-N16 of the Austrian Science Fund (FWF). 
  BA acknowledges funding by the contract ASI-INAF I/016/07/0, and ASI-INAF I/009/10/0.
  TL acknowledges support through the FWF-project P20046-N16.
  The United Kingdom Infrared Telescope is operated by the Joint Astronomy Centre on behalf of the Science 
  and Technology Facilities Council of the U.K.
  Science operations with PTI were conducted through the efforts of the 
  PTI Collaboration (http://pti.jpl.nasa.gov/ptimembers.html) and the 
  National Aeronautics and Space Administration (NASA) Exoplanet Science 
  Institute (NExScI), and we acknowledge the invaluable contributions of 
  our PTI colleagues. We particularly thank Kevin Rykoski for his 
  professional operation of PTI.  PTI was constructed with funds from the 
  Jet Propoulsion Laboratory, Caltech, as provided by NASA
  We acknowledge D.~V. Shulyak for computing the models for the standard stars, P.~Marigo and W.~Nowotny for the useful discussions,
  A.~Baier and S.~Sacuto for helping with Dusty Code, L.~Pavan for useful discussions on the reddening correction.
      
\end{acknowledgements}

\appendix
\section{Statistical approach for temperature determination}
\label{temp.app}
In order to assign an observed spectrum to one out of a sample of available models, a statistical approach is employed: each model 
represents a Gaussian-shaped sandheap the width of which (in terms of standard deviation) is determined by the point-to-point scatter $p$ of 
the spectrum. The height of the sandheaps can be adjusted by statistical weights. Thus the shape of the $n$th sandheap is given by the Gaussian probability density function
\begin{equation}
G\left( d_n\right) = \exp\left(\frac{d_n^2}{2p^2}\right)\: .
\end{equation}
The ``distance'' $d_n$ between the observed spectrum and the model is the rms deviation between the observed and the synthetic spectrum. 
The statistical analogy of the model assignment is to pick out a single grain of sand at the given distances from the synthetic spectra. 
Now the question is, ``What is the probability of the picked grain to belong to the $n$th sandheap?''

The total number of available grains at the selected position is determined by the sum of densities for all $N$ models,
\begin{equation}
\nu = \sum_{n=1}^{N}G\left( d_n\right)\: ,
\end{equation}
which represents our normalization condition. Thus the relative amount of grains belonging to model $n$ -- and hence the probability of th
e observed spectrum to represent this model -- evaluates to
\begin{equation}
\mathrm{Pr}_n = \frac{G\left( d_n\right)}{\nu}\: .
\end{equation}

In the context of our application, the only parameter of interest is the effective temperature, whence the temperature interval covered by the models is divided into a set of intervals $I_t$, $t=1,...,\tau$. Since the number of models belonging to such a temperature bin is not unique, the marginal probability density of temperature has to be corrected for this bias. In the sandheap metaphor, we would have to provide each temperature interval to be represented by a unique number of grains, and the probability of our spectrum to be assigned to the temperature interval $I_t$ evaluates to the weighted sum of individual model assignment probabilities
\begin{equation}
\mathrm{Pr}\left( I_t\right) = \frac{m_t^{-1}\sum_{n=1}^{m_t}\mathrm{Pr}_n\left( I_t\right)}{\sum_{i=1}^{\tau} m_i^{-1}\sum_{k=1}^{m_i}\mathrm{Pr}_k\left( I_i\right)}\: ,
\end{equation}
$m_t$ and denoting the number of models belonging to the interval $I_t$.

This normalisation is demonstrably valid, because the integral probability for all temperatures evaluates to
\begin{equation}
\sum_{t=1}^{\tau}\mathrm{Pr}\left( I_t\right) = \frac{\sum_{t=1}^{\tau}m_t^{-1}\sum_{n=1}^{m_t}\mathrm{Pr}_n\left( I_t\right)}{\sum_{i=1}^{\tau} m_i^{-1}\sum_{k=1}^{m_i}\mathrm{Pr}_k\left( I_i\right)} = 1\: .
\end{equation}

\listofobjects
\begin{longtable}{l l l l l l}
\caption{\label{vis.tab}Interferometric observations.}\\
       \hline
       \hline
       ID & UT Date & B & PA & $V \pm \sigma_V$ & Additional informations\\
       &&[m]&[deg]&\\
       \hline
\endfirsthead
\caption{continuated.}\\
\hline
\hline
      ID & UT Date & B & PA & $V \pm \sigma_V$ & Additional informations\\
       &&[m]&[deg]&\\
       \hline
\endhead
\hline
\endfoot
         CR~Gem  & 11/03/1999 & 101.49 & 80 & 0.324 $\pm$ 0.004& \\
       \hline
         HK~Lyr  & 05/20/1999 & 109.79 & 71 & 0.277 $\pm$ 0.004& \\
                 & 05/21/1999 & 109.78 & 78 & 0.277 $\pm$ 0.005& \\
                 & 05/22/2000 & 109.29 & 63 & 0.258 $\pm$ 0.005& \\
                 & 05/22/2000 & 109.33 & 64 & 0.265 $\pm$ 0.004& \\
                 & 05/22/2000 & 109.76 & 71 & 0.238 $\pm$ 0.010& \\
                 & 05/22/2000 & 109.77 & 72 & 0.244 $\pm$ 0.003& \\
                 & 05/22/2000 & 109.73 & 80 & 0.249 $\pm$ 0.003& \\
                 & 05/22/2000 & 109.72 & 81 & 0.242 $\pm$ 0.003& \\
\hline
        RV~Mon   & 11/04/1999 & 100.37 & 63 & 0.446 $\pm$ 0.015& \\
                 & 09/25/2008 & 108.68 & 52 & 0.305 $\pm$ 0.005& \\
                 & 09/25/2008 & 106.98 & 55 & 0.320 $\pm$ 0.004& \\
                 & 09/25/2008 & 104.79 & 57 & 0.349 $\pm$ 0.006& \\
\hline
           Z~Psc & 10/08/1995 & \,\,\,38.21 & $-$& 0.810 $\pm$ 0.123 & (1) \\ 
\hline
       DR~Ser    & 06/03/2000 & 106.29 & 55 & 0.387 $\pm$ 0.008& \\
                 & 06/19/2000 & 107.05 & 54 & 0.535 $\pm$ 0.024& (2)\\
                 & 06/19/2000 & 106.01 & 56 & 0.497 $\pm$ 0.035& (2)\\
                 & 06/19/2000 & 105.45 & 56 & 0.487 $\pm$ 0.032& (2)\\
                 & 06/19/2000 & 104.83 & 57 & 0.515 $\pm$ 0.024& (2)\\
                 & 06/20/2000 & 103.33 & 59 & 0.435 $\pm$ 0.013& \\
                 & 06/20/2000 & 102.19 & 60 & 0.429 $\pm$ 0.013& \\
                 & 04/18/2001 & 104.41 & 57 & 0.407 $\pm$ 0.010& \\
                 & 04/18/2001 & 104.20 & 58 & 0.415 $\pm$ 0.004& \\
                 & 04/18/2001 & 103.23 & 59 & 0.407 $\pm$ 0.003& \\
                 & 04/18/2001 & 103.06 & 59 & 0.413 $\pm$ 0.007& \\
                 & 04/18/2001 & 102.13 & 60 & 0.410 $\pm$ 0.005& \\
                 & 04/23/2001 & 106.14 & 55 & 0.418 $\pm$ 0.013& \\
                 & 04/23/2001 & 103.56 & 58 & 0.410 $\pm$ 0.012& \\
                 & 04/23/2001 & 100.63 & 62 & 0.429 $\pm$ 0.011& \\
                 & 04/23/2001 & \,\,\,99.18 & 64 &  0.435 $\pm$ 0.011& \\
                 & 04/23/2001 & \,\,\,97.73 & 66 &  0.440 $\pm$ 0.012& \\
                 & 04/29/2001 & 105.27 & 56 & 0.389 $\pm$ 0.006& \\
                 & 04/29/2001 & 105.01 & 57 & 0.391 $\pm$ 0.005& \\
                 & 04/29/2001 & 103.67 & 58 & 0.396 $\pm$ 0.005& \\
                 & 04/29/2001 & 102.84 & 59 & 0.406 $\pm$ 0.005& \\
                 & 04/29/2001 & 102.68 & 59 & 0.407 $\pm$ 0.008& \\
                 & 05/03/2001 & \,\,\,86.43 & 15 & 0.587 $\pm$ 0.014& \\
                 & 05/03/2001 & \,\,\,86.40 & 15 & 0.597 $\pm$ 0.015& \\
                 & 05/03/2001 & \,\,\,85.92 & 15 & 0.591 $\pm$ 0.019& \\
                 & 05/03/2001 & \,\,\,84.84 & 16 & 0.593 $\pm$ 0.018& \\
                 & 05/03/2001 & \,\,\,83.03 & 17 & 0.626 $\pm$ 0.023& \\
                 & 05/04/2001 & \,\,\,86.07 & 15 & 0.583 $\pm$ 0.009& \\
                 & 05/04/2001 & \,\,\,85.98 & 15 & 0.580 $\pm$ 0.009& \\
                 & 05/04/2001 & \,\,\,84.76 & 16 & 0.599 $\pm$ 0.010& \\
                 & 05/04/2001 & \,\,\,82.50 & 17 & 0.615 $\pm$ 0.007& \\
                 & 05/04/2001 & \,\,\,82.24 & 17 & 0.621 $\pm$ 0.011& \\
                 & 05/05/2001 & \,\,\,86.31 & 15 & 0.589 $\pm$ 0.006& \\
                 & 05/05/2001 & \,\,\,85.79 & 15 & 0.598 $\pm$ 0.007& \\
                 & 05/05/2001 & \,\,\,84.85 & 16 & 0.605 $\pm$ 0.005& \\
                 & 05/05/2001 & \,\,\,83.54 & 17 & 0.613 $\pm$ 0.008& \\
                 & 05/05/2001 & \,\,\,81.86 & 17 & 0.626 $\pm$ 0.006& \\
                 & 05/23/2001 & 104.92 & 57 & 0.402 $\pm$ 0.007& \\  
                 & 05/23/2001 & 103.80 & 58 & 0.416 $\pm$ 0.004& \\
                 & 06/21/2001 & 103.71 & 58 & 0.404 $\pm$ 0.006& \\
                 & 06/21/2001 & 102.06 & 60 & 0.420 $\pm$ 0.005& \\
                 & 06/21/2001 & 100.30 & 63 & 0.433 $\pm$ 0.005& \\
                 & 06/21/2001 & \,\,\,98.60 & 65 & 0.460 $\pm$ 0.009& \\
                 & 07/16/2001 & \,\,\,79.72 & 14 & 0.627 $\pm$ 0.012& \\
                 & 07/22/2001 & 105.69 & 56 & 0.402 $\pm$ 0.019& \\
                 & 07/22/2001 & 105.55 & 56 & 0.405 $\pm$ 0.019& \\
                 & 07/22/2001 & 104.73 & 57 & 0.393 $\pm$ 0.015& \\
                 & 07/22/2001 & 104.59 & 57 & 0.413 $\pm$ 0.016& \\
                 & 07/22/2001 & 101.55 & 61 & 0.427 $\pm$ 0.016& \\
                 & 07/22/2001 & 100.54 & 62 & 0.438 $\pm$ 0.014& \\
                 & 08/22/2001 & \,\,\,86.22 & 15 & 0.586 $\pm$ 0.022&\\
                 & 08/22/2001 & \,\,\,86.04 & 15 & 0.581 $\pm$ 0.020&\\
                 & 08/22/2001 & \,\,\,85.95 & 15 & 0.596 $\pm$ 0.022&\\
                 & 08/22/2001 & \,\,\,85.41 & 16 & 0.586 $\pm$ 0.026&\\
                 & 08/22/2001 & \,\,\,85.29 & 16 & 0.576 $\pm$ 0.020&\\
                 & 08/22/2001 & \,\,\,80.33 & 18 & 0.631 $\pm$ 0.028&\\
                 & 08/22/2001 & \,\,\,79.97 & 18 & 0.661 $\pm$ 0.055&\\
                 & 08/23/2001 & \,\,\,98.03 & 66 & 0.456 $\pm$ 0.011&\\
                 & 08/27/2001 & \,\,\,86.28 & 14 & 0.564 $\pm$ 0.005&\\
                 & 08/27/2001 & \,\,\,86.45 & 14 & 0.565 $\pm$ 0.003&\\
                 & 08/27/2001 & \,\,\,86.46 & 15 & 0.566 $\pm$ 0.004&\\
       \hline                                                      
\multicolumn{6}{l}{\tiny{(1) \citet{dyc96}}}   \\                                         
\multicolumn{6}{l}{\tiny{(2) spurious values, discarded of the computation} }\\                                           
\end{longtable}                                                    
                                                                   

\begin{thebibliography}{}

\bibitem[Abia \& Isern, 1997]{abi97}Abia, C., \& Isern, J. 1997, MNRAS, 289, 11  
\bibitem[Abia et al., 2002]{abi02}Abia, C., Dom\'inguez, I., Gallino, R., et al. 2002, ApJ, 579, 817
\bibitem[Abia et al., 2010]{abi10}Abia, C., Cunha, K., Cristallo, S., et al. 2010, \apj, 715, 94
\bibitem[Am\^{o}res \& Lep\'{i}ne, 2005]{amo05} Am\^{o}res, E.~B., \& L\'{e}pine, J.~R.~D. 2005, AJ, 130, 659 
\bibitem[Aringer et al., 1997]{ari97}Aringer, B., J\o rgensen, U.~G., \& Langhoff, S.~R. 1997, A\&A, 323, 202
\bibitem[Aringer, 2000]{ari00}Aringer, B., 2000, Ph.D. Thesis, University of Vienna
\bibitem[Aringer et al., 2009]{ari09}Aringer, B., Girardi, L., Nowotny, W., Marigo, P., \& Lederer, M.~T. 2009, A\&A, 503, 913
\bibitem[Aufdenberg et al., 2009]{auf09}Aufdenberg, J., Ridgway, S., \& White, R. 2009, in astro2010: The Astronomy and Astrophysics Decadal Survey, Science White Papers, astro2010S, 8
\bibitem[Bagnulo, 1996]{bag96}Bagnulo, S. 1996, Ph.D. Thesis, Belfast University
\bibitem[Beichman et al., 1988]{bei88}Beichman, C.~A., Neugebauer, G., Habing, H.~J., Clegg, P.~E., \and Chester, T. J. 1988, NASAR, 1190, 1 
\bibitem[Bergeat et al., 2002]{ber02}Bergeat, J., Knapik, A., \& Rutily, B. 2002, A\&A, 390, 967
\bibitem[Bergeat \& Chevallier, 2005]{ber05}Bergeat, J., \& Chevallier, L. 2005, A\&A, 429, 235
\bibitem[Boden et al., 1998]{bod98}Boden, A.~F., Colavita, M.~M., van Belle, G.~T., \& Shao, M.\ 1998, \procspie, 3350, 872
\bibitem[Boden et al., 1999]{bod99}Boden, A.~F., et al. 1999, \apj, 515, 356
\bibitem[Boffin et al., 1993]{bof93}Boffin, H.~M.~J., Abia, C., \& Rebolo, R. 1993, A\&AS, 102, 361  
\bibitem[Cavanagh et al., 2003]{cav03}Cavanagh, B., Hirst, P., Jenness, T., \& al. 2003, ASPC, 295, 237
\bibitem[Claussen et al., 1987]{cla87}Claussen, M.~J., Kleinmann, S.~G., Joyce, R.~R., \& Jura, M. 1987, ApJS, 65, 385 
\bibitem[Colavita et al., 1999]{col99} Colavita, M.~M., et al.\ 1999, \apj, 510, 505
\bibitem[Cutri et al., 2003]{cut03}Cutri, R.~M., Skrutskie, M.~F., van Dyk, S., et al. 2003, 
  in The IRSA 2MASS All-Sky Point Source Catalog, NASA/IPAC Infrared Science Archive
\bibitem[Dyck et al., 1996]{dyc96}Dyck, H.~M., van Belle, G.~T., \& Benson, J.~A. 1996, AJ, 112, 294
\bibitem[Epchtein et al., 1990]{epc90}Epchtein, N., Le Bertre, T., Lepine, J.~R.~D. 1990, A\&A, 227, 82
\bibitem[Eglitis \& Eglite, 1995]{egl95}Eglitis, I., \& Eglite, M. 1995, Ap\&SS, 229, 63
\bibitem[Erspamer \& North, 2003]{ers03}Erspamer, D., \& North, P. 2003, A\&A, 398,1121
\bibitem[Garc\'ia-Hernandez et al., 2007]{gar07}Garc\'ia-Hern\'andez, D.~A., Garc\'ia-Lario, P., Plez, B., et al., 2007, A\&A, 462, 711
\bibitem[Gautschy-Loidl et al., 2004]{gau04}Gautschy-Loidl, R., H\"ofner, S., J\o rgensen, U.~G., \& Hron, J. 2004, A\&A, 422, 289
\bibitem[Goorvitch \& Chackerian, 1994]{gor94}Goorvitch, D. \& Chackerian, Jr.~C. 1994, ApJS, 91, 483
\bibitem[Gustafsson et al., 1975]{gus75}Gustafsson, B., Bell, R.~A., Eriksson, K., \& Nordlund, A\&A 1975, A\&A, 42, 407
\bibitem[Gustafsson et al., 2008]{gus08}Gustafsson, B., Edvardsson, B., Eriksson, K., et al. 2008, A\&A, 486, 951
\bibitem[Harris et al., 2006]{har06}Harris, G.~J., Tennyson, J., Kaminsky, B.~M., Pavlenko, Y.~V., \& Jones, H.~R.~A. 2006, MNRAS, 367, 400
\bibitem[Hauschildt et al., 1999]{hau99} Hauschildt, P.~H., Allard, F., Ferguson, J., Baron, E. ,\& Alexander, D.~R. 1999, \apj 525, 871
\bibitem[Houk, 1963]{hou63}Houk, N. 1963, AJ, 68, 253 
\bibitem[Iben \& Renzini, 1983]{ibe83}Iben, I., \& Renzini, A. 1983, \araa, 21, 271
\bibitem[Ivezic \& Eliztur, 1997]{ive97}Ivezic, Z., \& Elitzur, M. 1997, MNRAS, 287, 799
\bibitem[J\o rgensen et al., 1992]{jor92}J\o rgensen, U.~G., Johnson, H.~R., \& Nordlund 1992, A\&A, 261, 263
\bibitem[J\o rgensen, 1997]{jor97}J\o rgensen, U.~G. 1997, 
  in Molecules in Astrophysics: Probes and Processes, ed. E.~F. van Dishoeck, IAU Symp. 178 (Kluwer, 441-456)
\bibitem[J\o rgensen et al., 1989]{jor89}J\o rgensen, U.~G., Almlof, J., Siegbahn, P.~E.~M. 1989, ApJ, 343, 554
\bibitem[J\o rgensen et al., 2000]{jor00}J\o rgensen, U.~G., Hron, J., \& Loidl, R. 2000, A\&A, 356, 253
\bibitem[Kerschbaum et al., 1996a]{ker96a}Kerschbaum, F., Lazaro, C., \& Habison, P. 1996a, A\&AS, 118, 397
\bibitem[Kerschbaum et al., 1996b]{ker96b}Kerschbaum, F., Olofsson, H., \& Hron, J. 1996b, A\&A, 311, 273 
\bibitem[Kervella et al., 2003]{ker03}Kervella, P., Th\'evenin, F., S\'egransan, D., et al. 2003, A\&A, 404, 1087
\bibitem[Kurucz, 1993]{kur93} Kurucz, R.~L. 1993, ASPC, 44, 87
\bibitem[Lambert et al., 1986]{lam86}Lambert, D.~L., Gustafsson, B., Eriksson, K., \& Hinkle, K.~H. 1986, ApJS, 62, 373
\bibitem[Lederer \& Aringer, 2009]{led09}Lederer, M.~T., \& Aringer, B. 2009, A\&A, 494, 403
\bibitem[Lester \& Neilson, 2008]{les08}Lester, J., \& Neilson, H. 2008, A\&A, 491, 633
\bibitem[Loidl et al., 2001]{loi01}Loidl, R., Lancon, A., \& Jorgensen, U.~G. 2001, A\&A, 371, 1065-1077
\bibitem[Marigo et al., 2008]{mar08}Marigo, P., Girardi, L., Bressan, A., \& et al. 2008, A\&A, 482, 883, 905
\bibitem[Masana et al., 2006]{mas06}Masana, E., Jordi, C., \& Ribas, I 2006, A\&A, 450, 735
\bibitem[Mattsson et al., 2010]{mat09}Mattsson, L., Wahlin, R., \& H\"ofner, S., 2010, A\&A, 509, 14 
\bibitem[Mozurkewich et al., 1991]{moz91}Mozurkewich, D., et al.\ 1991, \aj, 101, 2207
\bibitem[Neilson \& Lester, 2008]{nei08}Neilson, H.~R. \& Lester, J.~B. 2008, A\&A, 490, 807
\bibitem[Nowotny et al., 2005]{now05}Nowotny, W., Lebzelter, T., Hron, J., \& H\"ofner, S. 2005, A\&A, 437, 285
\bibitem[Nowotny et al., 2010]{now10}Nowotny, W., H\"ofner, S., \& Aringer, B. 2010, A\&A, 514, 35
\bibitem[Nowotny et al., 2011]{now11}Nowotny, W., Aringer, B., H\"ofner, S., \& Lederer, M.~T. 2011, A\&A, 529, 129
\bibitem[Ohnaka \& Tsuji, 1996]{ohn96}Ohnaka, K., \& Tsuji, T. 1996, A\&A, 310, 933
\bibitem[Paladini et al., 2009]{pal09}Paladini, C., Aringer, B., Hron, J., et al. 2009, A\&A, 501, 1073
\bibitem[Perryman et al., 1997]{per97}Perryman, M.~A.~C., Lindegren, L., Kovalevsky, J., Hoeg, E., Bastian, U., et al. 1997, A\&A, 323, 49
\bibitem[Pojmanski, 2002]{poj02}Pojmanski, G. 2002, Acta Astronomica, 52,397 
\bibitem[Querci et al., 1974]{que74}Querci, F., Querci, M., \& Tsuji, T. 1974, A\&A, 31, 265
\bibitem[Ramsay et al., 2004]{ram04}Ramsay Howat, S., Todd, S., Legget, S., et al. 2004, Proc. SPIE, 5492, 1160
\bibitem[Samus+, 2007-2009]{sam09}Samus, N.N., et al. 2009, General Catalogue of Variable Stars, yCat, 102025S  
\bibitem[Sch\"oier et al., 2005]{sch05}Sch\"oier, F.~L., Lindqvist, M., \& Olofsson, H. 2005, A\&A, 436, 633, 646
\bibitem[Shulyak et al., 2004]{shu04}Shulyak, D.~V., Tsymbal, V., Ryabchikova, T., St\"utz\, Ch., \& Weiss, W. W. 2004, \aap, 428, 993
\bibitem[Sloan et al., 1998]{slo98}Sloan, G.~C., Little-Marenin, I.~R., \& Price, S.~D. 1998, AJ, 115, 809-820
\bibitem[Solano \& Fernley, 1997]{sol97}Solano, E., \& Fernley, J. 1997, A\&AS, 122, 131
\bibitem[Tanaka et al., 2007]{tan07}Tanaka, M., Letip, A., Nishimaki, Y., \& et al. 2007, PASJ, 59, 939
\bibitem[Thompson et al., 2002]{tho02}Thompson, R.~R., Creech-Eakman, M.~J., \& van Belle, G.~T. 2002, ApJ, 577, 447  
\bibitem[Tsuji, 1986]{tsu86}Tsuji, T. 1986, \araa, 24, 89
\bibitem[Uttenthaler \& Lebzelter, 2010]{utt10}Uttenthaler, S., \& Lebzelter, T. 2010, A\&A, 510, 62 
\bibitem[van Belle \& van Belle, 2005]{van05} van Belle, G.~T., \& van Belle, G. 2005, \pasp, 117, 1263
\bibitem[van Belle et al., 2008]{van08} van Belle, G.~T., et al.\ 2008, \apjs, 176, 276
\bibitem[Vanture et al., 2007]{van07}Vanture, A.~ D., Smith, V.~V., Lutz, J., et al. 2007, \pasp, 119, 147
\bibitem[Verhoelst, 2005]{ver05}Verhoelst, T. 2005, PhD. Thesis, University of Leuven
\bibitem[Wittkowski et al., 2001]{wit01}Wittkowski, M., Hummel, C.~A., Johnston, K.~J., et al., 2001, A\&A, 377, 981
\bibitem[Wittkowski, 2004]{wit04a}Wittkowski, M. 2004, Proc. of the 37th Liege International Astrophysical Colloquium 
"Science Case for Next Generation Optical/Infrared Interferometric Facilities (the post VLTI era)" 
\bibitem[Wittkowski et al., 2004]{wit04b}Wittkowski, M., Aufdenberg, J.~P., \& Kervella, P. 2004, A\&A, 413,711
\bibitem[Wittkowski et al., 2006a]{wit06a}Wittkowski, M., Aufdenberg, J.~P., Driebe, T., et al. 2006a, A\&A, 460, 855 
\bibitem[Wittkowski et al., 2006b]{wit06b}Wittkowski, M., Hummel, C.~A., Aufderberg, J.~P., \& Roccatagliata, V. 2006b, A\&A, 460, 843  
\bibitem[Whitelock et al., 2008]{whi08}Whitelock, P.~A., Feast, M.~W.,\& van Leeuwen, F. 2008, MNRAS, 386, 313  
\end{thebibliography}
\end{document}